\documentclass{ws-ijmpa}
\usepackage{graphics} 
\usepackage{amssymb}

\def\Bbb{\bf}
\def\C{{\Bbb C}} \def\R{{\Bbb R}} \def\Z{{\Bbb Z}}

\def\id{\protect{{1 \kern-.28em {\rm l}}}}

\newcommand{\be}{\begin{equation}} \newcommand{\ee}{\end{equation}}
\newcommand{\bea}{\begin{eqnarray}} \newcommand{\eea}{\end{eqnarray}}
\newcommand{\beann}{\begin{eqnarray*}} \newcommand{\eeann}{\end{eqnarray*}}
\newcommand{\bfig}{\begin{figure}} \newcommand{\efig}{\end{figure}}
\newcommand{\nn}{\nonumber}
\newcommand{\ba}{\begin{array}}\newcommand{\ea}{\end{array}}

\begin{document}

\markboth{C.~I.~Lazaroiu}{D-brane categories}

%
%

\title{D-BRANE CATEGORIES}

\author{\footnotesize C.~I.~LAZAROIU}

\address{\it \footnotesize Institut f\"ur Physik\\Humboldt Universit\"at zu
Berlin\\ Newtonstrasse 15, 12489 Berlin,  Germany}

\maketitle


\begin{abstract}
This is an exposition of recent progress in the categorical approach
to D-brane physics. I discuss the physical underpinnings of the appearance of
homotopy categories and triangulated categories of D-branes from a string
field theoretic perspective, and with a focus on applications to homological
mirror symmetry. 
\end{abstract}


\vskip .6in

\section{Introduction}
\label{intro}

In his 1994 ICM lecture [\refcite{Kontsevich0}], 
M. Kontsevich suggested a vast categorical
extension of mirror symmetry, known as `homological
mirror symmetry'. This proposal put forward a very ambitious program aimed at
deriving mirror symmetry from a deeper correspondence between certain
categories associated with a mirror pair.  In physical language, Kontsevich's
proposal amounts to formulating mirror symmetry for open topological strings
in the presence of the most general D-branes, and then extract the usual
closed string formulation as a consequence.  While far from complete,
the basic ideas of this program have gradually penetrated a few areas of
inquiry, leading to valuable insights into D-brane dynamics. The purpose of
the present paper is to explain some results which have emerged from such
investigations.

Let us start with the broad expectations 
(which are in many ways rather conjectural). 
In a nutshell, Kontsevich's original proposal is as follows. If $X,Y$ is a
mirror pair of Calabi-Yau manifolds, then one considers two homological
objects associated with these spaces.  The object associated with $X$ is
$D^b(X)$, the bounded derived category of the Abelian category of coherent
sheaves on $X$. This is a triangulated category in the sense of Verdier.  The
object associated with $Y$ is a triangulated category $DFuk(Y)$, the derived
category of the Fukaya category of $Y$, whose objects are roughly a certain `quantum
version' of Lagrangian cycles in $Y$, and whose morphism spaces are given by
Floer homology. The rigorous construction of this category is now almost
complete [\refcite{Fukaya-book}, \refcite{Fukaya_mirror2}].
Given these two objects, Kontsevich's conjecture states:

{\bf Conjecture 0} {\em For $X$ and $Y$ a mirror Calabi-Yau pair, $D^b(X)$ and
$DFuk(Y)$ are equivalent as triangulated categories.}

The functor implementing this equivalence should be viewed as a
`derived' version of the mirror map.  There exists a `homotopical'
extension of this conjecture. For this, one notices that the two
categories $D^b(X)$ and $DFuk(Y)$ can be enhanced to certain minimal
$A_\infty$ categories, i.e. their associative products $r_2:Hom(A_2,A_3)\times Hom(A_1,A_2)\rightarrow 
Hom(A_1,A_3)$ can
be extended by higher products\footnote{Note that in this article we
follow the convention that morphisms compose `backwards'.}
$r_n:Hom(A_n, A_{n+1})\times \dots \times Hom(A_1, A_2)\rightarrow
Hom(A_1, A_{n+1})$ in order to obtain minimal $A_\infty$ categories
$D^b_\infty(X)$ and $D_\infty Fuk(X)$. The enhancement $D^b_\infty(X)$
results by considering a certain version of Massey products which is always well-defined and is 
induced by the deformation theory of complexes of locally free
sheaves [\refcite{Polishchuk_infty}]. It can also be realized more directly 
by constructing $D^b(X)$ as the derived category of 
a dG category and considering the minimal model of the category of twisted complexes appearing in that
construction. The $A_\infty$ enhancement $D_\infty Fuk(Y)$ results
directly from its construction as the derived category of an
$A_\infty$ category, by taking the minimal model of an $A_\infty$ category of twisted complexes 
which appears as an intermediate step in the construction. Then the `enhanced' version of homological
mirror symmetry states:

{\bf Conjecture 1} The two $A_\infty$ categories $D^b_\infty(X)$ and
$D_\infty Fuk(Y)$ are homotopy equivalent.

Note that Conjecture 1, if true, immediately implies Conjecture 0.

When $X$ and $Y$ are Calabi-Yau threefolds\footnote{The case of Calabi-Yau
manifolds of arbitrary dimension is less clear, since the formulation of
topological string field theory used in this article is appropriate only for 
threefolds.}, one way to understand the origin of this proposal is as
follows. Physically, one is interested in formulating mirror symmetry for open
strings. Following the approach pioneered by E. Witten, it suffices to
formulate such a correspondence at {\em topological} open string tree level,
i.e. in terms of open topological string amplitudes on a disk. The relevant
topological string theories are the open A model on $X$ and the open
B model on $Y$ [\refcite{Witten_nlsm}, \refcite{Witten_mirror}, 
\refcite{Witten_CS}]. 
One wants to consider open strings with `arbitrary'
boundary conditions, i.e. in the presence of the most general topological
branes. This can be achieved by formulating the problem in the language of string field
theory, starting with a distinguished set of boundary data (namely standard
D-branes of the B-model associated with holomorphic vector bundles on $X$,
respectively D-branes of the A-model  described by Lagrangians in $Y$ carrying Chan-Paton bundles). 
Then the `most general boundary data' can be recovered
indirectly by requiring closure of the resulting string field theory
under processes of D-brane composite formation. Modulo certain
(weak) assumptions, this recovers $D^b(X)$ and $DFuk(Y)$ as the required collections
of `generalized topological D-branes', and leads to extended string field theories defined
in terms of these categories.  Thus the weak form of the conjecture
(Conjecture zero) can be reformulated physically as follows:

{\bf Conjecture 0phys} The collection of topological D-branes of the B 
model with target space $X$ is equivalent to that of the A-model with target 
space $Y$. Moreover, this correspondence preserves the triangulated structure.

The $A_\infty$ enhancements $D^b_\infty(X)$ and $D_\infty Fuk(X)$ 
encode the (gauge-fixed) open string amplitudes on the disk, according to the
prescription:
\be
\langle \langle u_n\dots u_0\rangle\rangle_{tree} =
\langle u_n, r_n(u_{n-1}\dots u_0)\rangle~~, 
\ee
where the left hand side is the disk amplitude for scattering of the states
$u_0\in Hom(A_0, A_1)$ $\dots$  $u_{n-1}\in Hom(A_{n-1},A_{n})$ and $u_n\in
Hom(A_n,A_0)$ and $\langle ., .\rangle$ in the right hand side is a
bilinear pairing which plays the role of BPZ form for the topological open
string field theory under consideration.

The statement of Conjecture 1 can then be reformulated as follows:

{\bf Conjecture 1phys} For a mirror pair of Calabi-Yau {\em threefolds} $(X,Y)$, 
the `complete' open string field theories which incorporate the most general D-branes
of the topological A model on $X$ and the topological B-model on $Y$ should be equivalent as BV systems.

By a categorical extension of the results of [\refcite{Gaberdiel}], 
this conjecture implies homotopy equivalence of the associated $A_\infty$ categories, as
well as compatibility of this equivalence with the bilinear forms.

It is crucial for physical applications to consider the supplementary
information contained in the superconformal models of which the $A$ and $B$
models are twisted versions. This line of investigation has been pioneered
recently by M. Douglas and P. Aspinwall. In [\refcite{DA}], 
they have proposed a
way of selecting a subclass of B-type branes (an Abelian
subcategory of $D^b(X)$) which should be BPS saturated (and thus stable) when embedded in the
full superstring compactification on $X$ \footnote{A mathematical formulation
of this stability condition was recently given in [\refcite{Bridgeland_stab}].}.  
This fascinating
proposal has many interesting aspects, for example the relevant set of stable
objects must change with the stringy Kahler moduli of $X$, which gives an extremely general 
realization of the phenomenon of marginal stability known from supersymmetric field theories. It is currently not
entirely certain (from a physical point of view) 
what is the proper `mirror' of this stability condition,
which is expected to select a triangulated subcategory of the derived category of the 
Fukaya category
of $Y$ (though a natural conjecture follows from the work of 
[\refcite{Bridgeland_stab}] and was considered 
recently in [\refcite{Thomas_stab}]). 

Having recalled the broad conjectures, we now turn to discussing what is
known. This paper is
written from a physical perspective, but must assume familiarity with
certain mathematical notions which have been rarely used in the
physics literature. The string field theoretic point of
view used below is not the only possible approach to this
subject. However, it does provide a unifying perspective on the
origins of both of the categories $D^b_\infty(X)$ and
$D_\infty Fuk(Y)$, and gives a natural physical context for the
somewhat formidable theory of $A_\infty$ categories. For different
(though related) perspectives for the case of the B model, I refer the reader
to the work of M.~Douglas, P.~Aspinwall and collaborators 
[\refcite{DA}, \refcite{Aspinwall_D0}, \refcite{Aspinwall_massless}, 
\refcite{Aspinwall_sw}, \refcite{Douglas_Berenstein}, \refcite{Douglas_cat}].

\section{Generalized D-branes and open string field theory}
\label{osft}

One of the deepest recent insights into D-brane physics
is the re-consideration of the very concept of D-brane.  In the traditional approach,
one constructs D-branes by imposing boundary conditions on a string's
endpoints, and interprets the resulting string dynamics as a quantization of
the underlying space-time object carrying the boundary data. A crucial
observation is that such a description does not generally suffice, since it
cannot always account for all products of open string dynamics. For example,
the endpoint of tachyon condensation in a system of D-branes evolving in a
nontrivial background cannot always be described directly in this language. Thus
a complete description of open string dynamics in the presence of D-branes
requires a more general formulation. As originally suggested by A. Sen, this can 
be sought in the framework of open string field theory. In
this approach, one can give a conceptually clear description of D-brane
composite formation by using the language of category theory. More
precisely, one finds [\refcite{com1}, \refcite{com2}] that open (topological or
bosonic\footnote{In the bosonic case, there are functional-analytic
aspects which are not fully understood.  In this article, we shall
only need the topological case.}) string field theory in the presence
of D-branes (and formulated with Witten's choice of vertex 
[\refcite{Witten_SFT}]) can be
described in terms of a differential graded (dG) category endowed with
certain supplementary data, and that `closure' of this description
under processes of D-brane composite formation in the topological case
requires that the underlying dG category obey a certain
`quasiunitarity' constraint.  When applied to the string field theory
of open topological strings (the open A and B model), this
quasiunitarity condition requires that one extend any initial collection of 
D-branes by adding the totality of their composites. In this section, I give
a brief exposition of this analysis, which was carried out in
[\refcite{com1}, \refcite{com2}]. As we shall see below, the string field theory
approach recovers a construction originally proposed by Kontsevich
in his 1994 ICM lecture (and previously investigated in the
mathematical work of [\refcite{BK}]).  
By generalizing to an arbitrary choice of string field 
vertices along the lines of [\refcite{Gaberdiel}], 
this argument has an $A_\infty$
version which leads to a more general construction also 
sketched in Kontsevich's original talk and recently developed in 
[\refcite{Lefevre},\refcite{Fukaya_mirror2}]. 

Let us note that the open string field theory approach to the
classification of D-branes can be viewed as an `off-shell' extension (in the
sense of removing BRST closure constraints) of the on-shell discussion
given in [\refcite{Moore_top},\refcite{moore_K}] and 
[\refcite{top}] (though currently
this extension is only understood at open string tree level).  As
usual in physics, off-shell formulations allow one to extract more
information.

\subsection{Open string field theory with D-branes}
\label{axioms}

Categorical constructs enter D-brane physics through the very basic
structure involved in the description of open string dynamics.  One
way to understand their appearance is to extend the string-field
theoretic framework of [\refcite{Witten_SFT}] in order to incorporate
D-branes. For this, one treats D-branes as abstract objects, whose
worldsheet description is encoded in the algebraic data of string
products.  As explained in [\refcite{com1}], one can abstract the following
information, provided that
one uses Witten's choice of vertex in order to construct the string field action:

(1) A collection of objects $a$ (the `D-branes') together with complex
vector spaces $Hom(a,b)$ for any two objects $a$ and $b$.  These
describe the state
space of oriented strings stretching from $a$ to $b$, and carry an
integer-valued grading given by the ghost number of such states. The degree of
an element $u\in Hom(a,b)$ will be denoted by $|u|$.

(2) A collection of non-degenerate bilinear pairings $_{ab}\langle
.,.\rangle_{ba}:Hom(a,b)\times Hom(b,a)\rightarrow \C$ which describe the BPZ
forms

(3) A collection of trilinear maps $\langle \langle .,.,.\rangle\rangle :
Hom(c,a)\times Hom(b,c)\times Hom(a,b)\rightarrow \C$, which describe the
three-point functions on a disk with `boundary conditions' $a$,$b$,$c$. Upon
dualizing with respect to the BPZ forms, the same information be encoded in
bilinear compositions $Hom(b,c)\times Hom(a,b)\rightarrow Hom(a,c)$ defined
through: \be \langle \langle u,v,w\rangle\rangle =_{ca}\langle u,
vw\rangle_{ac}~~.  \ee 

(4) A collection of linear operators $d:Hom(a,b)\rightarrow Hom(a,b)$ of
degree $+1$, which square to zero. These are the worldsheet BRST operators acting on the
spaces of open string states.

This data is subject to certain compatibility constraints, which can be expressed
succinctly as follows:

(A) The objects $a$ form an (associative) category with morphism spaces
$Hom(a,b)$ and the composition defined as in point (3) above. When the
morphism spaces are endowed with the grading and with the BRST operators $d$, this becomes a
differential graded (dG) category. This means that $d$ squares to zero, acts as a derivation of
the composition operation, and that the latter has degree zero.

(B) The bilinear forms $\langle ., . \rangle$ are compatible with the
derivations $d$ and with the morphism compositions in the following sense:
\bea & & \langle du, v\rangle +(-1)^{|u|} \langle u, dv\rangle=0\nn\\ & &
\langle u, vw\rangle=\langle uv, w \rangle~~.  \eea

In the situation of interest below (namely a topological string field theory
defined on a Calabi-Yau threefold), the bilinear form must also satisfy the
selection rule: \be
\label{selrule}
\langle u, v \rangle =0~~{\rm unless}~~ |u|+|v|=3.  \ee

In the case of bosonic string field theory, one also has certain conjugation operations 
obeying natural compatibility conditions. We shall limit ourselves to topological string field
theories for what follows. Succinctly: 

{\em An open topological string field theory with D-branes is
described by a dG category ${\cal A}$ together with a collection of invariant
nondegenerate bilinear pairings (of total degree $-3$) on its morphism spaces.}

(For what follows, we shall also assume that ${\cal A}$ is endowed with direct sums. This can be easily achieved 
by adding such sums if necessary). 

Given this data, one can write a string field action as follows. The {\em
string field} is a degree one element $\phi$ of the {\em total boundary space}
${\cal H}=\oplus_{a,b}{Hom(a,b)}$. The string
field action has the form: \be
\label{S}
S(\phi)=\frac{1}{2}\langle \phi, d\phi\rangle +\frac{1}{3}\langle \phi,
\phi\phi\rangle \ee where the {\em total BPZ form} $\langle . , .\rangle$ is
obtained from $_{ab}\langle \cdot, \cdot \rangle_{ba}$ by summing over the components $Hom(a,b)$.
Here $d$ is the differential induced on ${\cal H}$ and we let $\bullet$ (usually denoted by juxtaposition) 
be the composition induced on this space.

\subsection{Vacua and D-brane composites}
\label{composites}
String field vacua are obtained by extremizing $S(\phi)$, which gives the
equation: \be
\label{MC}
d\phi +\frac{1}{2} [\phi,\phi]=0~~, \ee where $[.,.]$ is the graded commutator
built from the associative composition $\bullet$ on ${\cal H}$. This can be recognized as the
Maurer-Cartan equation governing deformations of the differential graded (dG)
Lie algebra $({\cal H}, d,[.,.])$ ($=$the commutator algebra of the
differential graded associative algebra $({\cal H}, d, \bullet )$). In terms of
the total boundary data $d$, ${\cal H}$ and $\bullet$, the vacua are described
in the manner familiar from open string field theory without D-branes. The new
element is the category structure provided by the D-brane labels $a$, which is
compatible with the total boundary data in sense that $d$ preserves the spaces
$Hom(a,b)$, the boundary product $\bullet$ maps $Hom(b,c)\times Hom(a,b)$ into
$Hom(a,c)$ and the bilinear form $\langle \cdot , \cdot \rangle$ vanishes on
$Hom(a,b)\times Hom(c,d)$ unless $b=c$ and $a=d$.

Equations (\ref{MC}) have the trivial solution $\phi=0$ (this reflects the
fact that we consider a background-dependent formulation of open string field
theory, expanded around a given vacuum).  A
nontrivial solution $\phi=\alpha$ of (\ref{MC}) defines a new vacuum, which can be
viewed as a translation of this distinguished background. The total BRST
operator for the expansion around the new background has the form: \be d_\alpha
=d +[\alpha, \cdot]~~.  \ee Equation (\ref{MC}) can be viewed as the tadpole
cancellation condition for the translated background and is equivalent with
the requirement $d_\alpha^2=0$. The crucial observation is that $d_\alpha$ need not
be compatible with the original category structure. This means that {\em
shifting the string vacuum changes the D-brane content of the theory}, an
observation which allows one to give an abstract description of D-brane
dynamics. Namely, {\em the D-brane content described by a vacuum $\alpha$ is
encoded in a category structure ${\cal A}_\alpha$ which is compatible with
the new BRST operator $d_\alpha$}. This category describes the
D-brane content obtained by expanding the theory around the vacuum $\alpha$, and
was constructed explicitly in [\refcite{com1}], where it was called the {\em
collapsed category}. In that reference, it was also showed that ${\cal
A}_\alpha$ obeys all axioms of open string field theory with D-branes, with
respect to bilinear forms and compositions constructed naturally by shifting
the string vacuum.

Shifting the vacuum by a solution $\phi=\alpha=\oplus_{a,b}{\alpha_{ab}}$ of (\ref{MC})
amounts to condensing spacetime fields associated with the components
$\alpha_{ab}\in Hom(a,b)$. This leads to the formation of D-brane composites
out of the branes described by the objects $a$ for which there exists an
object $b$ with $\alpha_{ab}\neq 0$ or $\alpha_{ba}\neq 0$ (the branes lying in the support $S$ of
$\alpha$). In the simplest case when $S$ is `connected' (in graph-theoretic 
sense explained in [\refcite{com1}]), 
the collapsed category ${\cal A}_\alpha$ is obtained upon replacing
these D-branes (the objects lying in the support of $\alpha$) with a new object
$*$, identified with the resulting D-brane composite. The new morphism spaces,
compositions and bilinear forms are constructed accordingly. The result ${\cal
A}_\alpha$ of this construction is a dG category carrying bilinear forms subject
to the axioms Section \ref{axioms}.  Hence (topological) tree-level open string dynamics in the
presence of D-branes can be described by passage to the collapsed string field
theory associated to whatever open string background one cares to
condense. 

\subsection{Closure under composite formation and the quasiunitary cover}
\label{closure}

A complete description of D-brane dynamics must be closed under
formation of D-brane composites. This means that the theory ${\cal A}_\alpha$
obtained after shifting the string vacuum should be a `sub-theory' of the
original string field theory, in the sense that all possible D-brane composites should 
already be considered as objects in the original theory. 
This requirement was formalized in [\refcite{com1}],
as the condition that the collapsed category ${\cal A}_\alpha$ is dG-equivalent
with a subcategory of the original dG category ${\cal A}$ 
(the bilinear forms can also be matched). In [\refcite{com1}], an open
string field theory with D-branes which satisfies this completeness
requirement was called `quasi-unitary'. As explained in [\refcite{com1}], 
any open
string field theory with D-branes ${\cal A}$ can be extended to a quasiunitary
theory which `contains' it in an appropriate sense. In fact, there exists a
minimal extension of this type, the so-called `quasi-unitary cover' $c({\cal
A})$ of ${\cal A}$, which is constructed as follows:

(1) The objects of $c({\cal A})$ are {\em generalized complexes of degree one} over ${\cal
A}$. A generalized complex of degree one is a finite sequence $(a_j)_j$ of objects of
${\cal A}$, together with degree one morphisms $q_{ij}\in Hom^1_{\cal
A}(a_i,b_j)$ such that $q:=\oplus_{i,j}{q_{ij}}$ satisfies the Maurer-Cartan
equation (\ref{MC}):
\be
\label{twisted_complex}
dq_{ij}+\sum_{k}{q_{kj}q_{ik}}=0~~.
\ee
 
(2) Given two generalized complexes $A:=(a_k, q_{ij})$ and $B:=(b_k, q'_{ij})$,
the space $Hom_{c({\cal A})}(A,B)$ is given by the direct sum
$\oplus_{i,j}{Hom_{\cal A}(a_i,b_j)}$, with the induced grading. The
differential on this space is defined by: 
\be
du=\oplus_{i, j}{\left[
du_{ij}+\sum_{k}{q'_{kj}u_{ik}}-
\sum_{l}{(-1)^{|u_{lj}|}u_{lj}q_{il}}\right]}
\ee
for $u=\oplus_{i, j}{u_{ij}}\in Hom(A,B)$, with $u_{ij}\in Hom(a_i,b_j)$.

(3) Given a third degree one generalized complex $C=(c_k, q''_{ij})$, 
the composition of morphisms $Hom(B,C)\times Hom(A,B) \rightarrow Hom(A,C)$ 
is given by:
\be
uv=\oplus_{i, k}{\left[\sum_{j}{u_{jk}v_{ij}}\right]}~~,
\ee
where $u=\oplus_{j, k}{u_{jk}}\in Hom(B,C), v=\oplus_{i,j}
{v_{ij}}\in Hom(A, B)$, with $u_{jk}\in Hom(b_j,c_k)$ and $v_{ij}\in Hom(a_i,b_j)$.

(4) Finally, one has natural bilinear forms on the morphism spaces
of $c({\cal A})$ and it is easy to check that the axioms of Section \ref{axioms} are satisfied (see [\refcite{com1}]).

This physically-motivated construction is entirely general and relies only on the
most basic data which specify an open (topological) string field theory with D-branes. 
As we shall see below, it is realized explicitly for the open B-model, as well as in a sector of the open A-model.
Moreover, a homotopy version of it is realized through Fukaya's approach to topological A-type branes, 
leading to Kontsevich's proposal [\refcite{Kontsevich0}] 
for the construction of the derived category of Fukaya's category.

\subsection{Topological A/B strings and graded topological D-branes}
\label{graded_branes}

If one considers the open A or B model on a Calabi-Yau space $M$, then one
must take into account some supplementary data which enter a complete
description of boundary sectors. This
is easiest to see in the context of the A-model, for which a basic D-brane
is described by a Lagrangian cycle of $M$ (which carries 
Chan-Paton data and has vanishing Maslow index etc). As already pointed out by Kontsevich
in his 1994 lecture [\refcite{Kontsevich0}] 
and elaborated by Seidel [\refcite{Seidel_graded}], a
well-defined $\Z$-grading on the state spaces of open strings stretching
between two Lagrangians requires that one specifies their `relative' grading 
(this is necessary even in the case of special Lagrangians, which 
describe D-branes of the untwisted model). A similar grading can be
introduced for the basic topological D-branes of the B-model (which are described by
holomorphic vector bundles). This grading is specified by a discrete choice and 
arises naturally in at the topological level 
[\refcite{Kontsevich0}, 
\refcite{Seidel_graded}, \refcite{Douglas_cat}, \refcite{com2}, 
\refcite{Diac}, \refcite{sc}]. 
Its effect is to shift the worldsheet $U(1)$ charge of a string stretching from a 
brane $a$ to a brane $b$ according to the formula:
\be
\label{topshift}
|\phi_{ba}|\rightarrow |\phi_{ba}| +grade(b)-grade(a)~~.
\ee

More recently, M. Douglas introduced a different type
of grade (which is real-valued) [\refcite{Douglas_cat}]. 
This is relevant for D-branes of the full {\em
super}string compactification, to which the topological models are related by
twisting. The grade introduced by M. Douglas is related to the
space-time central charge of the D-brane, when the latter is embedded into the full 
superstring model.  This real grading should be
treated as supplementary data which allows one to make contact with
superstring physics, by picking out those topological D-branes which are
conjectured to correspond to BPS saturated branes of the associated superstring compactification (as we
discuss later).  For the moment, we focus on the topological aspects of
Kontsevich's conjecture.

The effect of introducing the integer grading at the topological string field
level was first discussed in [\refcite{com2}] 
\footnote{A discussion based on that
of [\refcite{com1}] and [\refcite{com2}] 
was later given in [\refcite{Diac}] for the B
model. See also [\refcite{sc}] for the effect of this on a certain sector 
of the $A$-model.  }.  
As explained in [\refcite{com2}], the combination
of the string field theory analysis of the previous section with the
supplementary data described by the integer grading makes immediate contact
with the older work of Bondal and Kapranov [\refcite{BK}] 
while providing a clear
physical justification for a construction already proposed in Kontsevich's
original lecture [\refcite{Kontsevich0}]. The argument is 
based on the string field theoretic analysis of D-brane composite formation
which was sketched above. This argument works for any graded string field
theory, and in particular can be applied for both the $A$ and $B$ models.  For
the $A$-model, a formulation in terms of $A_\infty$ categories is more
practical in view of the work of [\refcite{Fukaya-book}], 
as we will discuss later.

\subsection{Graded open string field theory}
\label{graded_sft}

We say that an open string field theory with D-branes is {\em (integer)
graded} if the underlying dG category ${\cal A}$ is endowed with shift
functors $[n]$ ($n\in \Z$) compatible with the bilinear forms. Given a theory as in Subsection \ref{axioms}, 
one can
consider the minimal graded theory containing it, the so-called {\em
shift-completion} of the original theory. This is obtained by adding objects
$a[n]$ for each integer $n$ and each object $a$ of the original theory, and
adding morphism spaces, differentials, compositions and bilinear forms in the
obvious manner dictated by shift-invariance. In particular, the original
objects $a$ can be identified with the new objects $a[0]$, and the original dG
category ${\cal A}$ becomes a full subcategory of its shift-completion ${\tilde {\cal A}}$. It is
easy to check that ${\tilde {\cal A}}$ satisfies the axioms of Section
\ref{axioms} when endowed with the bilinear forms constructed in the obvious manner.

In the shift-completed theory ${\tilde {\cal A}}$, one has the basic relation:
\be 
Hom(a[m],b[n])=Hom(a,b)[m-n]~~, 
\ee 
where
$Hom(a,b)[p]^k:=Hom^{k+p}(a,b)$. Combining with the construction discussed
above, one finds that the quasiunitary cover $c({\tilde {\cal A}})$ of the
shift-completed theory is described through data familiar from the
work of [\refcite{BK}]. Namely, applying the description of
generalized complexes given in the previous subsection shows that a degree one
generalized complex over ${\tilde {\cal A}}$ is given by a finite sequence of
objects $a_i[n_i]$ of ${\tilde {\cal A}}$ (with $a_i$ objects of ${\cal A}$),
together with morphisms $q_{ij}\in Hom_{\tilde {\cal
A}}^{1}(a_i[n_i],a_j[n_j])=Hom^{1+n_i-n_j}(a_i,a_j)$, subject to the
Maurer-Cartan equation (\ref{twisted_complex}) with respect to the composition
on ${\tilde {\cal A}}$.

Such objects were originally considered in the mathematics literature 
[\refcite{BK}],
where they were called {\em twisted complexes} over ${\cal A}$.  In the
language of [\refcite{BK}], 
the twisted complexes predicted by string field theory
are {\em two-sided}, i.e.  the morphisms $q_{ij}$ are not required to vanish
for $i\geq j$ (this was pointed out in [\refcite{com2}]). If one does impose
this condition (which is useful for technical reasons), then one obtains the
concept of {\em one-sided} twisted complexes. The dG category $c({\tilde {\cal
A}})$ of two-sided twisted complexes over ${\cal A}$ will be denoted by
$tw({\cal A})$. As originally showed in [\refcite{BK}], 
one-sided twisted complexes
also form a dG category, which we shall denote by $tw^+({\cal A})$. In
conclusion, the quasiunitary cover of the shift-completed theory $c({\tilde
{\cal A}})$ coincides with the category $tw({\cal A})$ of {\em two-sided}
twisted complexes\footnote{Provided that one uses {\em sequences} in the
construction of twisted complexes, rather than sets of objects, as discussed
in [\refcite{com1},\refcite{com2}].} : 
\be c({\tilde {\cal A}})=tw({\cal A})~~.  \ee This
relation was pointed out in [\refcite{com2}] upon building on the work of
[\refcite{com1}] (see also [\refcite{Diac}] and [\refcite{sc}]).

As discussed in [\refcite{BK}], the associative category $H^0(tw^+({\cal A}))$
obtained by taking the cohomology of $tw^+({\cal A})$ in degree zero is
triangulated. This category will be called the {\em derived category of the dG
category ${\cal A}$} and denoted by $D({\cal A})$. Passage to this category
allows one to identify topological D-branes up to quasi-isomorphisms, a
process which is permitted by the Batalin-Vilkovisky formalism. Since
$tw^+({\cal A})$ is a subcategory of $tw({\cal A})$, it immediately follows
that $D({\cal A})=H^0(tw^+({\cal A}))$ is a subcategory of the associative
category $H^0(tw({\cal A}))$ predicted by string field theory. In general,
there is no clear reason to believe that $D({\cal A})$ and $H^0(tw({\cal A}))$
are equivalent, though one can come close to finding an equivalence between
them.

For various results about dG categories, their enhanced triangulated
categories and the associated theory of functors I refer the reader to
[\refcite{BK}, \refcite{Keller_dg}, \refcite{Drinfeld}].  
The case of two-sided twisted complexes is
largely understudied.

\section{The $A_\infty$ structure induced by gauge-fixing and the deformation potential}
\label{def}

Returning to the general framework of Section \ref{axioms}, let us consider
the string field action (\ref{S}). This has the infinitesimal
gauge-invariance: \be
\label{gauge_infin}
\phi\rightarrow \phi+d\beta+[\phi,\beta]~~, \ee for generators $\beta\in {\cal
H}^0$ (the finite gauge transformations are easily obtained by exponentiating
(\ref{gauge_infin})). We consider the problem of building a tree-level
effective potential for fluctuations of $\phi$ in the vicinity of a solution
of the equations of motion (\ref{MC}). This was discussed in 
[\refcite{superpot}],
with the following result. Fluctuations which are not pure gauge can be
separated explicitly provided that the theory under consideration is endowed
with supplementary data (technically, what is needed is a so-called {\em
cohomological splitting} of the underlying $dG$ category). The most direct
physical interpretation arises in a theory satisfying (\ref{selrule}),
provided that the cohomological splitting is induced by an antilinear
involution $c:{\cal H}\rightarrow {\cal H}$, which is compatible with the
category structure and satisfies $|cu|=3-|u|$. Under this assumption, one can
define a nondegenerate Hermitian metric on ${\cal H}$ through $h(u,v):=\langle
cu, v \rangle$ and partially fix the gauge invariance (\ref{gauge_infin}) by
imposing the condition $d^\dagger\phi=0$, where $d^\dagger$ is the Hermitian
conjugate of $d$ with respect to $h$. The degree one component of the space
$K=ker d\cap ker d^\dagger$ describes massless fluctuations around the vacuum
and can be identified with the degree one cohomology $H^1_d({\cal H})$.  One
has a natural propagator $U=\frac{1}{d}\pi_d$, where $\pi_d$ is the
orthoprojector on $im d$ (with respect to the metric $h$). When expanding
around the solution $\phi=0$ of (\ref{MC}), the massive modes can be
integrated out to produce an effective potential $W$ for the massless modes
$u\in K^1$, whose tree level piece $W_{tree}$ can be easily described by
diagrams involving the propagator $U$. The result is: \be
\label{superpot}
W_{tree}(u)=\sum_{n\geq 3}{\frac{1}{n+1}(-1)^{n(n-1)/2} \langle \langle
u,\dots ,u\rangle \rangle^{(n+1)}_{tree}}~~, \ee where: \be
\label{amplitudes}
\langle \langle u_0\dots u_n\rangle \rangle^{(n+1)}_{tree}= \langle u_0,
r_{n}(u_1\dots u_n)\rangle~~ \ee for $u_j\in K$. Here $r_n:K^{\otimes
n}\rightarrow K$ are multilinear maps which can be described in terms of the
propagator and the associative composition on ${\cal H}$. As discussed in
[\refcite{superpot}], the tree level amplitudes satisfy certain cyclicity
constraints, while the maps $(r_n)_{n\geq 2}$ form a minimal $A_\infty$
algebra. In a theory with multiple D-branes, the operator $c$ must be
compatible with the category structure, in which case the products $r_n$
define a minimal $A_\infty$ category with the same objects and morphism spaces
as the homology category $H^*({\cal A})$. Moreover, this minimal $A_\infty$
category is quasi-isomorphic with the original dG category ${\cal A}$ (by
results of [\refcite{Fukaya_mirror2}], 
such a quasi-isomorphism is automatically a
homotopy equivalence). As discussed in [\refcite{superpot}], 
extremizing $W_{tree}$
over $K^1$ and modding out residual complex symmetries is equivalent with
solving the Maurer-Cartan equations (\ref{MC}) and dividing through the full
gauge symmetry of the original theory. This gives an alternative description of
the local moduli space, and shows that $W_{tree}$ encodes the obstructions to
infinitesimal deformations of the vacuum sitting at $\phi=0$\footnote{The fact
that an effective potential of this type should govern vacuum deformations of
pure Chern-Simons theory in three dimensions was suggested a while ago by
E. Witten [\refcite{Witten_CS}].  The potential (\ref{superpot}) implements this
idea in the more general case of open string field theory, whose action is
formally of Chern-Simons type.  More recent work on such deformation
potentials can be found in [\refcite{Kajiura}]. }.

The $A_\infty$ algebra described by $(K, \{r_n\})$ is the so-called {\em
minimal model} [\refcite{Kadeishvili}] of the dG algebra 
$({\cal H},d,\bullet)$ (a
similar terminology applies at the category level). The physical construction
of the minimal model which follows as in [\refcite{superpot}] by integrating out
the massive modes in perturbation theory coincides with a mathematical
construction discussed for example in [\refcite{Kontsevich_Soibelman}]. 
This gives
a physics-inspired proof of the fact that a dG algebra (and, more generally,
an $A_\infty$ algebra [\refcite{Kontsevich_Soibelman}]) always admits a minimal
model, i.e. a homotopically-equivalent $A_\infty$ algebra whose first product
$r_1$ vanishes.  A similar result holds for $A_\infty$ categories. Hence the
minimal model theorem for $A_\infty$ algebras and categories is a mathematical
reflection of the existence of an effective potential for massless modes. As
explained in [\refcite{superpot}], 
this effective potential is holomorphic (when
convergent) with respect to coordinates defined by a linear basis of $K^1$,
and, in the general case of multiple boundary sectors, it can be viewed as a
potential for the low energy dynamics of arbitrary systems of topological
D-branes.  In the context of the string field theory obtained by twisting a
superstring compactification on a Calabi-Yau threefold, it can be viewed as a
generalization of the `D-brane superpotential' of [\refcite{Kachru_superpot}].  The
potential $W_{tree}$ is defined on the `virtual tangent space' at a point
$\phi$ of the moduli space (in the discussion above, we chose $\phi=0$).  The
former is the space of {\em linearized} deformations described by
$H^1_{d_\phi}({\cal H})\approx K^1$, which is typically of higher dimension
than the dimension of the (highest local component of the) moduli space. We
note that generalized D-branes are typically heavily obstructed, hence the
deformation potential rarely vanishes for a nontrivial object of a topological
D-brane category.

The construction performed in [\refcite{superpot}] depends on the existence of a
conjugation operator $c$ with certain properties (a formulation which was
chosen there due to its physical character). However, it immediately follows
from the results of [\refcite{superpot}] that the main property of this
superpotential (namely that it encodes obstructions to the deformations of a
given vacuum) remains valid under more general assumptions. In fact, one can
replace the string scattering products $r_n$ by the products of any
homotopy-equivalent model of the original dG category ${\cal A}$ in order to
describe such deformations. This always leads to the categorical homotopy
version of the Maurer-Cartan equations which is discussed, for example, in
[\refcite{Fukaya_rev}]. In particular, the minimal model theorem can be applied for
any cohomological splitting, and thus such a `low energy' description of
deformations can be extracted via more abstract means.

It is clear from the discussion above that the $A_\infty$ structure is crucial
both physically and mathematically, even if one formulates the underlying open
string field theory by using Witten's vertex.  Physically, this structure
encodes the data of the D-brane superpotential of 
[\refcite{Kachru_superpot}] at an
extremely general level. Thus the enhanced version of homological mirror
symmetry (Conjecture 1 in the introduction) expresses matching of tree-level
open string scattering amplitudes between mirror systems of D-branes or,
equivalently, matching of the associated D-brane superpotentials.

\section{Topological B-type branes and the derived category}
\label{Bmodel}

The string field theoretic framework of the previous section can be realized
very explicitly for the open sector of the topological B model originally
introduced by E. Witten 
[\refcite{Witten_nlsm}, \refcite{Witten_mirror}, \refcite{Witten_CS}].  Given a
Calabi-Yau threefold $X$, the simplest topological D-branes of the associated
B-model are described by holomorphic vector bundles over $X$. For any two such
bundles $E_1, E_2$, localization shows that the (off-shell) state space of
topological open strings stretching from $E_1$ to $E_2$ can be identified with
$\Omega^{0,*}(X, E_1^*\otimes E_2)$, while the boundary BRST operator $d$
becomes the Dolbeault differential ${\overline \partial}$ coupled to the
bundle $E_1^*\otimes E_2$. Thus our original class of objects is $Vect(X)$,
the collection of all holomorphic vector bundles on $X$, while the morphism
spaces are given by $Hom(E_1, E_2):=\Omega^{0,*}(X,E_1^*\otimes E_2)$ with the
obvious grading.  An immediate generalization of [\refcite{Witten_CS}] 
shows that
the string product $Hom(E_2,E_3)\times Hom(E_1, E_2)\rightarrow Hom(E_1, E_3)$
is given by the wedge product of bundle-valued forms (which includes
composition in the bundle directions). Finally, the BPZ form on $Hom(E_2,
E_1)\times Hom(E_1, E_2)$ is given by: \be
\label{B_pairing}
\langle u, v\rangle =\int_{X}{\Omega\wedge tr (u\wedge v)} ~ ~.  \ee This data
gives a dG category ${\cal A}=Vect_{dg}$, endowed with the nondegenerate
bilinear forms (\ref{B_pairing}) which are easily seen to obey all axioms of
Section \ref{osft}.

As explained in [\refcite{Douglas_cat}], this picture must be completed by
introducing an integer-valued grade\footnote{Recall that there also exists a
different type of grade, which is real-valued, and which is relevant for the
untwisted model. This will be discussed later.} for each brane $E\in Vect(X)$,
which can be described abstractly by adding all of the formal translates
$E[n]$. Following the general discussion of the previous section, this leads
to the shift-completed string field theory ${\tilde {\cal A}}$, which was
originally considered in [\refcite{com2}]. 
As discussed in [\refcite{Diac}], this theory
admits an interesting sector which is obtained upon restricting to an object
$E$ together with its degree translates $E[n]$.  In this case, the string
field action can be written in terms of a graded superconnection ${\cal B}$ of
total degree $(0,1)$ on the graded bundle ${\bf E}=\oplus_{n}{E[n]}$: \be
\label{BCS}
S:=\int_{X}{\Omega\wedge str\left[\frac{1}{2}{\cal B}{\bar D}{\cal
B}+\frac{1}{3}{\cal B}^3\right]}~~.  \ee This can be viewed as a background
independent formulation of our string field theory, when restricted to such a
sector.  The original formulation used above arises upon choosing a background
superconnection ${\cal B}_0$ and expanding ${\cal B}={\cal B}_0+\phi$, where
$\phi$ plays the role of the string field of Section \ref{osft}.  Expanding
$\phi$ into its components along $Hom(E[m], E[n])$ recovers the
category-theoretic description, as it applies to this particular sector. In
particular, the equations of motion of (\ref{BCS}) become the Maurer-Cartan
equations for $\phi$, which recover twisted complexes upon expanding
$\phi$ in its components.

Returning to the general formulation of the previous section, one can
now pass to the category $c({\tilde {\cal A}})=H^0(tw({\cal A}))$.  If one
restricts to one-sided 
twisted complexes for which $q_{ij}$ vanishes unless
$n_i <n_j$ (a technical step whose physical
justification in topological string field theory is unclear),  
then one is left with usual complexes of vector bundles since these are 
the only one-sided twisted complexes which satisfy that condition.
Because of this, one 
finds that $D^b(X)$ can be identified with a subcategory of 
$H^0(tw({\cal A}))$, a relation which follows easily from the 
results of [\refcite{BK}] (see [\refcite{com2}, \refcite{Diac}, 
\refcite{Aspinwall_Lawrence}]).  Via this identification, the $A_\infty$
structure which governs the deformation potential of Section \ref{def}
extends to give the $A_\infty$ enhancement $D^b_\infty(X)$ of $D^b(X)$
entering Conjecture 1 of the introduction [\refcite{Polishchuk_infty}].
An immediate consequence of the description of topological 
B-branes as objects of the derived 
category is a general realization of open string state spaces as
Ext groups. Some aspects of how this description 
arises from the nonlinear sigma model were recently studied in 
[\refcite{Sharpe}].

\section{Topological A branes and the Fukaya category}
\label{Amodel}

This section gives a brief discussion of Fukaya's category, including
an explanation of its physical origins. I also discuss work on a
certain sector of this category, which leads to a graded version of
Chern-Simons field theory, and describe some of the results obtained
in that direction.

Since a systematic presentation of the
mathematical theory of $A_\infty$ categories is outside the scope of this article, I refer the reader to
the paper [\refcite{Fukaya_mirror2}] and the thesis [\refcite{Lefevre}] 
(see [\refcite{Keller_intro}] for an introduction), where
the basic theory of such categories is developed in some detail. The
role played by $A_\infty$ algebras in open string field theory (with one boundary sector) 
was discussed in [\refcite{Gaberdiel}]. It parallels the well-known
fact [\refcite{Witten_Zwiebach}, \refcite{Voronov}]
that $L_\infty$ algebras appear naturally in closed (bosonic or topological)
string field theory, as the algebraic constraints satisfied by tree-level
products as a consequence of Ward identities. For open strings, $A_\infty$
categories (as opposed to $A_\infty$ algebras) arise in the framework of
[\refcite{Gaberdiel}] 
simply by introducing a collection of D-branes. As well-known
from the work of B.~Zwiebach (see, for example, [\refcite{Zwiebach_oc}]), 
the
construction of a (bosonic or topological) string field theory requires a
choice of vertices. For (bosonic or topological) open string field theory, the
most widely-used choice is that employed by E. Witten in 
[\refcite{Witten_SFT}], which requires only one (triple)
vertex and thus leads to a single string product $r_2$. Algebraically, this
choice leads to a differential graded category, as discussed in Section \ref{osft} (where $r_2$ was 
identified -- up to sign factors -- with the composition $\bullet$).
On the other hand, a general choice of vertices leads to an infinity of string products
which are subject to $A_\infty$ constraints, thus requiring the algebraic framework of $A_\infty$ algebras 
[\refcite{Gaberdiel}], which  
is appropriate for a single boundary sector. In the presence of a collection of D-branes, this 
framework must be further generalized to that of $A_\infty$ categories. It turns out that Fukaya's
construction uses a general choice of vertices, which is why $A_\infty$ structures
are relevant in that situation. We also note that by results of 
[\refcite{Fukaya_mirror2}], an 
$A_\infty$ category admits a so-called `anti-minimal' model, i.e. it is homotopy-equivalent 
with a dG category. Therefore, one can in principle used the dG framework of 
[\refcite{Witten_SFT}, \refcite{com1}] 
throughout. However, this is not always technically advantageous, and the homotopy 
equivalence required to recover the anti-minimal description can be extremely complicated. In the 
case of Fukaya's category, for example, such a homotopy equivalence will involve summation over all
disk instanton contributions. 

Let us mention that one of the main results of the theory of $A_\infty$ algebras is that 
a quasi-isomorphism between such objects is the same as a homotopy equivalence 
(see [\refcite{Fukaya-book}]). A similar result is proved in 
[\refcite{Fukaya_mirror2}] for 
$A_\infty$ categories. This shows that the homological and homotopical classification of such 
objects coincide.

\subsection{A physical overview}

As compared to the B-model, the case of topological A-branes is considerably
more difficult. The main reason is the presence of worldsheet instanton
corrections for the open A-model, which leads to a series of phenomena studied
in detail in 
[\refcite{Floer1}, \refcite{Floer2}, \refcite{Oh}, 
\refcite{Fukaya_mirror1}, \refcite{Fukaya-book}, \refcite{Fukaya_mirror2}]. 
Since one attempts to build an open
string field theory, it suffices to study tree level instantons, which in this
case amount to (pseudo)-holomorphic maps from a disk to the target space,
subject to so-called Lagrangian boundary conditions [\refcite{Witten_CS}].

Let us first explain the main physical ideas. The open $A$-model with target
space a Calabi-Yau threefold $Y$ was introduced in [\refcite{Witten_CS}]. 
This is a
cohomological field theory governing maps $\phi:\Sigma\rightarrow Y$ where
$\Sigma$ is a Riemann surface with boundary.  As showed in 
[\refcite{Witten_CS}],
the model admits Lagrangian boundary conditions, a class of boundary conditions which respect 
the BRST symmetry. The associated topological branes are described by
pairs $(L,E)$ where $L$ is a Lagrangian submanifold\footnote{It is
also possible to consider the more general case of Lagrangian
immersions, but we shall not discuss that here.} of $Y$ and $E$ is a
vector bundle on $L$ endowed\footnote{This is a slight extension of
the work of [\refcite{Witten_CS}], where only bundles carrying a flat
connection (B=0) where considered.}  with a connection whose curvature
equals $2\pi i B|_L$ ($B$ is the B-field on $Y$). Remember that a
submanifold $L$ of $Y$ is called {\em Lagrangian} if its real
dimension equals the complex dimension of $Y$ and if $\omega|_L=0$,
where $\omega$ is the Kahler form of $Y$. Physically, $E$ is the
bundle of Chan-Paton factors. As we shall see in the next subsection,
one should add supplementary data, namely a {\em grading} 
[\refcite{Seidel_graded}] and a `relative spin' structure 
[\refcite{Fukaya-book}] for $L$.

As in the closed case, the open A-model suffers instanton corrections,
with worldsheet instantons described by the condition that $\phi$ is
holomorphic.  One can consider strings whose endpoints end on a single
brane $(L,E)$, or strings stretching between two D-branes $(L_1, E_1)$
and $(L_2, E_2)$. In the first case, open string states localize on
elements of $\Omega^*(L, End(E))$, while in the second case they are
described by elements of $\oplus_{p\in L_1\cap L_2}{Hom(E_1|_p,
E_2|_p)}$, where we assume for simplicity that $L_1$ and $L_2$ have
transverse intersection. These spaces describe off-shell open strings
states in these boundary sectors. To describe {\em physical} string
states in the first case is considerably more complicated (due to the
effect of instanton corrections), as we shall see in a moment. The
final result will be that such on-shell (physical) state spaces are
described by a very general version of Floer homology developed in
[\refcite{Fukaya-book}, \refcite{Fukaya_mirror2}].

The most general tree-level situation is to specify a finite
collection of topological A-branes $(L_j, E_j)$ ($j=1\dots n$) and
study scattering amplitudes on the disk.  Namely, one fixes some
points $z_j$ on the boundary of the disk $D$ such that
$\phi(z_j)=p_j\in L_j\cap L_{j+1}$ (where $L_{n+1}:=L_1)$ as well as
mutually distinct points $w_i^{(j)}$ on $\partial D$ (with $i=1\dots
l_j$) such that $\phi(w_i^{(j)})=q_i^{(j)}\in L_j$, and scatters
elements of $Hom(E_j|_{p_j}, E_j|_{p_{j+1}})$ and
$\Omega^*(L_j, End(E_j))$. Here we assumed that
the ordering of $z_j$ on the boundary agrees with the orientation of
the disk.  The scattering amplitude will involve integration over the
moduli space of the boundary-punctured disk (the configuration space
of boundary points $z_j$ and $w^{(j)}_i$ divided by the obvious
$PSL(2,\R)$ action). Through localization, the associated path
integral gives an integral over the moduli space of holomorphic maps
$\phi$ from $D$ to $Y$, subject to the conditions $\phi(D_j)\subset
L_j$ and $\phi(z_j)=p_j$ (one must also integrate over $q_i^{(j)}$ in
$L_j$ so these points do not give extra-constraints). Here $D_j$ is
the segment along the disk's boundary lying between the points $p_j$
and $p_{j+1}$. This moduli space is of course non-compact, and a
proper definition of the integral requires finding an appropriate
compactification and defining a fundamental cycle. The latter requires
perturbation of the complex structure of $Y$ to an almost complex
structure, which is physically allowed since the open A-model
continues to be well-defined and cohomological in this more general
situation. Such technical difficulties are familiar from the study of
Gromov-Witten invariants, but they are more severe in our case since
the moduli spaces involved are not complex and thus cannot be
approached with the methods of algebraic geometry.  Instead, an
analytic approach is required. This was developed in 
[\refcite{Floer1}, \refcite{Floer2}, \refcite{Fukaya_infty}, 
\refcite{Fukaya_mirror1}, \refcite{Oh}, \refcite{Fukaya-book}, 
\refcite{Fukaya_mirror2}] 
and its results will be briefly recalled below. As we shall see, one must use a certain 
modification of the description of topological A-branes through pairs $(L,E)$ (namely one 
must use `graded Lagrangian submanifolds' $L$ 
[\refcite{Seidel_graded}] which carry a so-called 
`relative spin structure').

The main difficulty arising in the study of open A-type strings is the
fact that worldsheet instanton corrections induce tadpole
contributions which may displace the original string background. This
phenomenon was originally discovered by K. Fukaya 
[\refcite{Fukaya_mirror1}]
and later also noticed in [\refcite{Kachru_superpot}].  It arises from
strings ending on a brane $(L,E)$ (namely when $\phi(\partial
D)\subset L$, without boundary insertions) and signals the fact that
the description of the brane through the pair $(L,E)$ is generally
only valid semiclassically (by which we mean in the absence of
worldsheet instanton corrections). In [\refcite{Fukaya-book}, 
\refcite{Fukaya_rev}],
this phenomenon is described as an `obstruction' to the existence of
Floer homology. To make sense of the underlying theory (in the sector
where strings stretch from $L$ to itself) one must find a shift of the
string vacuum which cancels the tadpole (of course, it is possible
that such a deformation of $(L,E)$ does not exist, in which case the
associated topological brane will be destabilized by worldsheet
instanton corrections).  The process of shifting the string background
can be described most elegantly in the framework of string field
theory. Since the approach of Fukaya uses string amplitudes directly,
we must employ the general formalism of open string field theory,
which results from that of [\refcite{Witten_SFT}] by generalizing the
choice of vertex. Recall from [\refcite{Gaberdiel}] that:

{\em A (topological) open string field theory with a single
boundary sector, defined with a general choice of vertices and expanded
around a background which satisfies the equations of motion, is
described by a set of string products which satisfy the constraints
of an $A_\infty$ algebra, together with a compatible and
nondegenerate bilinear form (the BPZ form).}

In the conventions used in this paper, the products $r_n$ have degree $2-n$ and 
the bilinear form has  degree $-3$\footnote{This is related to the conventions of [\refcite{Fukaya_mirror2}] by suspension, namely Fukaya's products $m_n$ are given by
$r_n=s m_n(s^{\otimes -n})$ where $s$ is the suspension map. 
This introduces certain sign prefactors in our formulae. 
We prefer to work with $r_n$ since they are more directly related to the case of dG algebras. }.
The string field action has the form:
\be
S(\phi)=\sum_{n\geq 0}{\frac{1}{n+1}(-1)^{n(n-1)/2}\langle \phi, r_n(\phi^{\otimes n})\rangle}~~, 
\ee
where $\phi$ is a degree one element of the boundary space. 
Compatibility of the bilinear form means that the string products
satisfy certain cyclicity conditions, which are given in
[\refcite{Gaberdiel}].

The statement above applies only if the string field theory is
expanded around a background which satisfies the equations of
motion. It can be shown [\refcite{CIL_sft}] that expanding the string field
theory around a general background (which need not satisfy the
equations of motion) leads to a so-called {\em weak $A_\infty$
algebra}, namely the obvious  generalizing of an $A_\infty$ algebra obtained by 
allowing for a supplementary product $r_0:\C\rightarrow {\cal H}$ (this mathematical concept was introduced 
in [\refcite{Fukaya_mirror1}]). The
products $r_n$ obey a slight generalization of the $A_\infty$
constraints familiar from the case of standard (`strong') $A_\infty$
algebras, for example: 
\bea
r_1(r_0(1))&=&0~~\\
r_2(r_0(1),u)\pm r_2(u, r_0(1))\pm r_1(r_1(u))&=&0\nn
\eea 
and higher relations.  In the
context of string field theory, $r_0(1)$ describes the contribution of
the tadpole, which induces a linear term in the string field action.
As in [\refcite{Gaberdiel}], the product $r_1$ plays the role of worldsheet
BRST charge; it fails to square to zero due to 
the presence of tadpoles.  Shifting the string background is achieved
by a translation $\phi\rightarrow \phi+\alpha$ of the string field, which
leads to a new set of products $r'_n$ obtained by substituting this shift in the action: 
\be
r'_n(u_1\dots u_n):=\sum_{j_0\dots j_n \geq 0}{(-1)^{N} r_{n+j_0+\dots
+j_n}(\alpha^{\otimes{j_0}}, u_1, \alpha^{\otimes j_1}, u_2\dots u_n, \alpha^{\otimes
j_n})}~~.  
\ee
where $(-1)^N$ is a sign factor obtained by suspension.  
The condition that the shifted background satisfies the
equations of motion is equivalent with vanishing of the tadpole
contribution $r'_0$ around the new background; this automatically
assures that $(r'_n)_{n\geq 1}$ satisfy the (usual) $A_\infty$
constraints, and in particular the shifted worldsheet BRST charge
$r'_1$ squares to zero.  The condition $r'_0=0$ gives the `weak homotopy
Maurer-Cartan equation': 
\be
\label{whmc}
\sum_{n\geq 0}{(-1)^{n(n-1)/2}r_n(\alpha^{\otimes n})}=0 
\ee 
This equation admits a
natural gauge equivalence, which allows\footnote{Properly speaking, it allows one to define 
a (local) deformation functor. To build a moduli space one must perform a Kuranishi analysis.} 
for building a local moduli space as
explained in [\refcite{Fukaya_rev}]. It is of course possible that
(\ref{whmc}) does not have any solutions, in which case the original
background does not admit a continuous deformation to a string
vacuum. If (\ref{whmc}) does admit solutions, then the shifted
backgrounds can be described abstractly by the pair (original
background, solution $\alpha$ of (\ref{whmc})). This is the physical
interpretation of the procedure used in [\refcite{Fukaya-book}].

As in the case of associative string field theory (which is described by
Witten's choice of vertex [\refcite{Witten_SFT}]), 
the more general description of
[\refcite{Gaberdiel}] admits a categorical extension. Such an extension is
appropriate (and required) when one has more than one boundary sector, namely
when writing the string field action in the presence of more than one
D-brane. This extension [\refcite{CIL_sft}] is formulated as follows:

{\em A (topological) open string field theory with D-branes, constructed with a general
choice of vertices and expanded around a background which satisfies the
equations of motion, is described by an $A_\infty$ category ${\cal A}$ together
with compatible and nondegenerate bilinear forms (of degree $-3$) on its morphism spaces.}

Compatibility of the bilinear form means that the $A_\infty$ products of the
category satisfy cyclicity relations with respect to these forms, which are
the obvious categorical generalization of the cyclicity conditions given in
[\refcite{Gaberdiel}] (written there with conventions different from ours).  I will
not write the relevant formulae here, because they are quite
complicated. Instead, the reader is referred to [\refcite{Fukaya_mirror2}] for a
clear mathematical treatment of such data (except for the bilinear forms and
cyclicity, which are easily recovered by adapting the work of
[\refcite{Gaberdiel}]). When expanding around a background which fails to satisfy
the equations of motion, one must again include zeroth order products $r_0$
which describe the tadpoles correcting the D-branes. The result is that the
$A_\infty$ category is replaced by a weak $A_\infty$ category, the obvious
generalization of a weak $A_\infty$ algebra. Finally, the procedure of
shifting the string vacuum has an obvious analogue in the presence of
D-branes, and the relevant formulae are obtained from those given above by
expanding in boundary sectors. Once again, we shall assume that ${\cal A}$ is
endowed with direct sums.

Granted that one expands around a string vacuum, one can again consider the
problem of D-brane composite formation. Arguments very similar to those of
[\refcite{com1}] show that closure under such processes generally requires an
enlargement of the underlying $A_\infty$ category, which is obtained by a
`homotopy' analogue of the discussion of Section \ref{composites}.  This shows
that the original $A_\infty$ category ${\cal A}$ must be extended to its
quasi-unitary cover $c({\cal A})$, which is an $A_\infty$ category whose
objects are two-sided generalized complexes over ${\cal A}$ (defined by the
obvious $A_\infty$ version of
the generalized complexes discussed in Section \ref{osft}). 

For a {\em graded} string field theory, generalized complexes become two-sided
twisted complexes (the two-sided version of the twisted complexes used in
[\refcite{Lefevre}, \refcite{Kontsevich0}, \refcite{Fukaya_mirror2}]).  
One can again restrict to
one-sided complexes, a technically advantageous step whose physical
justification is unclear at the topological string level. With this
restriction, one obtains an $A_\infty$ category $tw^+({\cal A})$, whose
construction was originally suggested in [\refcite{Kontsevich0}] 
and carried out in
detail in [\refcite{Fukaya_mirror2}, \refcite{Lefevre}]. The (zeroth) homology category
$D{\cal A}$ of $tw^+({\cal A})$ (where the homology is taken with respect to
the products $r_1$ of $tw^+({\cal A})$) is a triangulated category as showed,
for example, in [\refcite{Fukaya_mirror2}, \refcite{Lefevre}]. This is the derived category
of the $A_\infty$ category ${\cal A}$.
 
Summing up, the physical description of the work of [\refcite{Fukaya-book},
\refcite{Fukaya_mirror2}] (at least for the case of Calabi-Yau threefolds) is as follows:

(1) One builds string amplitudes on the disk associated to scattering of
    open string states between finite systems of `graded' topological
    A-branes described by pairs $(L, E)$ with $L$ a graded Lagrangian
    submanifold carrying a relative spin
    structure.

(2) One uses these to build an open string field theory in the general
    sense of [\refcite{Gaberdiel}] (and including multiple boundary
    sectors). Due to use of graded Lagrangians, this string field
    theory will be graded.

(3) Because of the presence of tadpoles correcting the D-branes, one must shift the original D-brane
background to a true string vacuum. This requires solving the equations of
motion of the theory, which are encoded in the category-theoretic version of
(\ref{whmc}). When solutions exist, then each choice of solution specifies a
deformation of the original set of D-branes, which is induced by worldsheet
instanton corrections. Choose such a deformation $\alpha$ of the D-brane
background and shift to the associated string vacuum to obtain an $A_\infty$
category ${\cal A}_\alpha$ endowed with shift functors. This is the Fukaya
category $Fuk(Y)$\footnote{It is possible that more than one solution $\alpha$
of the tadpole cancellation conditions exists.  In that case,
one can construct a different version of Fukaya's category by including all
possible quantum deformations of the original collection of topological 
D-branes. This is related to the
theory of deformations of $A_\infty$ categories, currently being developed by
Kontsevich and collaborators.}.

(4) To implement closure of the physical description under formation of
D-brane composites, one must pass to the quasi-unitary cover of $Fuk(Y)$,
which is achieved by introducing twisted complexes. Restricting to one-sided
twisted complexes, one obtains an $A_\infty$ category $tw^+(Fuk(Y))$.  Its
zeroth cohomology $H^0(tw^+(Fuk(Y))$ is the derived category $DFuk(Y)$ of the
Fukaya category.

\subsection{Graded A-type branes}

As originally pointed out in [\refcite{Seidel_graded}], a well-defined $\Z$-grading
(as opposed to $\Z_2$-grading) on Floer homology requires that one add certain
discrete data to the description of topological A-branes originally proposed
in [\refcite{Witten_CS}]. Since Floer homology describes the on-shell state space
of open A-strings (which is $\Z$-graded by worldsheet $U(1)$ charge), this
fact clearly has physical relevance, which was explained in
[\refcite{Douglas_cat}]. As in the case of the B-model, this becomes important when
one considers at least {\em two } D-branes, and is necessary in order to fix
an ambiguity in the description of the worldsheet $U(1)$ charge. The ambiguity
is specified by an integer (which, following [\refcite{Douglas_cat}], we shall call
the {\em grade} of the brane\footnote{This is the {\em integer} grade 
relevant at topological sigma-model level, which in particular specifies 
a choice of branch for the real-valued grade of [\refcite{Douglas_cat}].}), 
which in the worldsheet language of the
untwisted theory describes the winding number of the boson which appears by
bosonizing the worldsheet $U(1)$ current [\refcite{Douglas_cat}]. Since this is a
conformal field theory argument, it will only be valid if the branes under
consideration are described by {\em special} Lagrangian cycles. It is known
[\refcite{Fukaya_rev}] that the Maslow index $\mu:\pi_2(Y,L)\rightarrow \Z$ of a
special Lagrangian cycle is identically zero, a condition which is believed
[\refcite{Thomas_Yau}] to be the symplectic-topology equivalent of the special
Lagrangian constraint.  As we recall below, a Lagrangian cycle can be graded
only if its Maslow index vanishes.

For any point $p$ in $Y$, consider the {\em oriented Lagrangian Grassmannian}
${\cal L}(T_pY)$, i.e. the Grassmannian of all oriented Lagrangian linear
subspaces of $T_pY$.  This is a space whose fundamental group equals $\Z$.
Varying $p$ inside $Y$, we obtain a bundle ${\cal L}(Y)$.  Since $Y$ is a
Calabi-Yau manifold, we have $c_1(Y)=0$ which implies [\refcite{Seidel_graded},
\refcite{Fukaya_mirror2}] that there exists a cover ${\tilde {\cal L}}(Y)$ of this bundle whose
restriction to each point $p$ of $Y$ can be identified with the universal
cover of ${\cal L}(T_pY)$. Choosing such a cover ${\tilde {\cal L}}(Y)$ makes
the underlying symplectic space of $Y$ into a {\em graded symplectic
manifold} (and we shall fix such a cover below). Given a Lagrangian submanifold $L$ in $Y$, we have the so-called
{\em Gauss map}, which is the section of ${\cal L}(Y)$ over $L$ given by: \be
\sigma(p):=T_pY~~{\rm for}~~p\in L~~.  \ee

A {\em grading} of $L$ is a lift of this map to the cover ${\tilde {\cal
L}}(Y)$. As explained in [\refcite{Fukaya_mirror2}], such a lift exists only if $L$
has vanishing Maslow index, in which case there is a countable number of such
lifts. A {\em graded Lagrangian submanifold} of $Y$ is simply a Lagrangian
submanifold $L$ endowed with such a lift.  We shall denote a graded
Lagrangian by the same letter $L$,the grading being understood. We then use
$L[n]$ to denote the same cycle $L$, but with the grading shifted by $n$
(though the action of $\Z$ on ${\tilde {\cal L}}(Y)$). When $L$ is special
Lagrangian, there is a canonical choice of grading which defines an `origin'
in the set of all gradings of $L$ [\refcite{Seidel_graded}]; 
this is why the grading of a special
Lagrangian can be simply viewed as an integer.  As for the B-model, the effect
of the grading is to shift the worldsheet $U(1)$ charge of string states
according to relation (\ref{topshift}). The complete description of topological
A-brane configurations requires the specification of this data, which is
essential in the construction of Fukaya's category.

\subsection{Relative spin structure}

As explained in [\refcite{Fukaya-book}, \refcite{Fukaya_mirror2}], 
the construction of string field
products also requires that our Lagrangian submanifolds $L$ admit a so-called
'relative spin structure' -- this is necessary in order to give an
orientation to the moduli spaces of open string instantons. To define this
data, one must fix a class $t\in H^2(Y,\Z_2)$. Then there is is a unique rank
2 real vector bundle $V$ on $Y$ of Stiefel-Whitney classes $w_1(V)=0, w_2(V)=t$
(remember that we assume $dim_\C Y=3$). We shall also assume that all our
Lagrangian submanifolds are oriented. Then a {\em relative spin structure} 
[\refcite{Fukaya-book}] on
$L$ with respect to $t$ is a spin structure on the restriction of the bundle
$V|_L\oplus TL$ to the two-skeleton of $L$. Such a structure exists if $L$ is oriented and
$w_2(TL)=t|_L$, in which case one says that $L$ is {\em relatively spin}. We
shall use the letter $L$ to denote a (gradable and relatively spin)
Lagrangian cycle together with its grading and choice of relative spin
structure.

\subsection{A (very) brief review of Fukaya's results}

As explained in [\refcite{Fukaya-book}, \refcite{Fukaya_mirror2}], 
the objects of Fukaya's category
are oriented Lagrangian submanifolds of $Y$ which are both graded and endowed
with a relative spin structure (for technical reasons, the construction
proceeds through a choice of a countable family of such cycles).  To construct
the string products $r_n$, one must consider families of objects (described by pairs 
formed by such a cycle and a Chan-Paton bundle) $a_j$
($j=1\dots n+1$) and scatter strings as explained at the beginning of this
section.  However, because objects are graded, it is now possible for example
that $a_{j+1}=a_j[1]$ in such a sequence.  Thus one effect of the grading is
that one must consider disk amplitudes in which one of the strings which are
scattered stretches between an object $a$ and the object $b=a[n]$, which
geometrically has the same underlying cycle $L$ but with grading shifted by
$n$. In this case, the grading on the state space of strings stretching from
$a$ to $b$ must be shifted according to relation (\ref{topshift}): \be
[Hom(E_a, E_b)\otimes \Lambda^* (L)]^k=Hom(E_a, E_b)\otimes \Lambda^{k-n}
(L)~~.  \ee An example of this situation is showed in figure 1.

\hskip 1.0 in
\begin{center} 
\scalebox{0.5}{\begin{picture}(0,0)%
\includegraphics{pair.pstex}%
\end{picture}%
\setlength{\unitlength}{4144sp}%
\begingroup\makeatletter\ifx\SetFigFont\undefined%
\gdef\SetFigFont#1#2#3#4#5{%
  \reset@font\fontsize{#1}{#2pt}%
  \fontfamily{#3}\fontseries{#4}\fontshape{#5}%
  \selectfont}%
\fi\endgroup%
\begin{picture}(4393,4123)(783,-4769)
\put(2656,-1366){\makebox(0,0)[lb]{\smash{\SetFigFont{17}{20.4}{\familydefault}{\mddefault}{\updefault}
$a$
}}}
\put(3106,-2581){\makebox(0,0)[lb]{\smash{\SetFigFont{17}{20.4}{\familydefault}{\mddefault}{\updefault}
$Hom(a,a)$
}}}
\put(4096,-871){\makebox(0,0)[lb]{\smash{\SetFigFont{17}{20.4}{\familydefault}{\mddefault}{\updefault}
$Hom(a,b)$
}}}
\put(2701,-961){\makebox(0,0)[lb]{\smash{\SetFigFont{17}{20.4}{\familydefault}{\mddefault}{\updefault}
$b$
}}}
\put(5176,-1681){\makebox(0,0)[lb]{\smash{\SetFigFont{17}{20.4}{\familydefault}{\mddefault}{\updefault}
$Hom(b,b)$
}}}
\put(5131,-3166){\makebox(0,0)[lb]{\smash{\SetFigFont{17}{20.4}{\familydefault}{\mddefault}{\updefault}
$Hom(b,a)$
}}}
\end{picture}
}
\end{center}
\begin{center} 
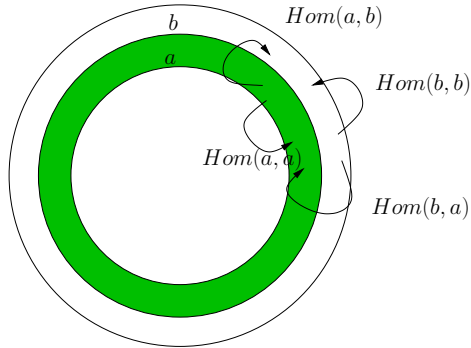
Figure 1. {\footnotesize Boundary sectors for a pair of graded D-branes
wrapping the same special Lagrangian cycle. The two D-branes $a$ and $b$ are
thickened out for clarity, though their (classical) \ thickness is zero.}
\label{pair}
\end{center}

As an extreme case of this, one can consider the sector described by a brane
$(L,E)$ and all branes obtained from it by shifting the grade. This is the
case considered in the work of [\refcite{sc}, \refcite{bvf}, \refcite{gf1}, 
\refcite{gf2}], which we discuss in
the next subsection. As explained above, such D-brane sectors are {\em
required} in the construction of Fukaya's category (see Section 4 of Fukaya's
paper [\refcite{Fukaya_mirror2}] for the discussion of these sectors). 
Similarly, they
are required in the open B-model, as explained in the
previous section.  Without including such sectors, one cannot obtain a
triangulated category in the end (since one needs the existence of shift
functors for that purpose).

The construction of string products for an entire (countable) collection of
cycles is given in [\refcite{Fukaya-book}, \refcite{Fukaya_mirror2}], and I will refer the
reader to those references for the (very technical) details. As described
above, the result is essentially a weak $A_\infty$ category. In fact, this
concept is not entirely appropriate [\refcite{Fukaya_mirror2}, 
\refcite{Fukaya-book}],
since one must deal with convergence issues for the disk instanton expansion,
which requires one to work over the `positive' Novikov ring $\Lambda_{0,nov}$
and introduce an energy filtration. Because of this, the technically correct
version of weak $A_\infty$ category employed in [\refcite{Fukaya_mirror2},
\refcite{Fukaya-book}] 
is a so-called {\em filtered $A_\infty$ category}. I refer the
reader to  Ref. \refcite{Fukaya_mirror2} for a detailed description of this
notion. Then the main result of 
[\refcite{Fukaya-book}, \refcite{Fukaya_mirror2}] is as
follows\footnote{Strictly speaking, the result quoted below has been proved
(as of the time of writing) only for the case of Chan-Paton data described by
line bundles.}:

\begin{theorem}
Fix $t\in H^2(Y,\Z_2)$ and a
cover ${\tilde {\cal L}}(Y)$ of ${\cal L}(Y)$.  Also fix an appropriate
countable collection of graded Lagrangian submanifolds $L$ of $X$, each
endowed with a relative spin structure (with respect to $t$) and with a
complex vector bundle $E$ carrying a unitary connection whose curvature equals
$2\pi i B|_{L}$. Call these {\em objects} and denote them by $a_j$.  Assume
that for any two such objects, the underlying Lagrangian submanifolds either
coincide or intersect transversely. Also assume that this collection of
objects is closed under the action of shifts of grading. Then there exists a
filtered $A_\infty$ category ${\cal A}$ with objects $a_j$, which encodes the
disk amplitudes of the underlying A-model in the presence of the topological
D-branes described by this countable set of objects.
\end{theorem}

The morphism spaces $Hom(a, b)$ of this category are defined as follows. Let
$a=(L_a, E_a, {\tilde \sigma}_a, rst(a))$ and $b=(L_b, E_b, {\tilde \sigma}_b,
rst(b))$ be two objects, where ${\tilde \sigma}$ are the associated lifts of
the Gausss map (which specify the grading) and $rst$ are the relative spin
structures. Then:

(1) If $L_a$ and $L_b$ intersect transversely, then 
one has $Hom(a,b)=\oplus_{p\in
L_a\cap L_b}{Hom(E_a|_p, E_b|_p)\otimes \Lambda_{0,nov}}$, with the grading
induced by the so-called {\em absolute Maslow index} (see
[\refcite{Seidel_graded}, \refcite{Fukaya_mirror2}]).

(2) If $L_b=L_a[n]$, then $Hom(L_a,L_b)=C(a,b){\hat \otimes}\Lambda_{0,nov}$,
where $C(a,b)$ is a certain countably-generated subcomplex of the complex
$W^{-\infty}(Hom(E_a,E_b)\otimes \Lambda^*(L))$ of distribution-valued forms,
with the grading induced by the following twisted grading on
$Hom(E_a,E_b)\otimes \Lambda^*(L)$: \be [Hom(E_a,E_b)\otimes
\Lambda^*(L)]^g:=Hom(E_a,E_b)\otimes \Lambda^{g-n}(L)~~.  \ee In the
definition of $Hom(a,b)$, the symbol ${\hat \otimes}$ denotes completion
of the tensor product with respect to the metric induced by the energy
filtration.  Consideration of distribution-valued forms and the restriction to
the subcomplex $C(a,b)$ is necessary for technical reasons which are explained
in [\refcite{Fukaya-book}].

As outlined above, the next step is to shift the string vacuum in order to
eliminate the zero-th products $r_0$, thus obtaining a (true) $A_\infty$
category $Fuk(Y)={\cal A}_\alpha$ defined over the Novikov ring.  Then one
constructs the category $tw^+(Fuk(Y))$ by considering (one-sided) twisted
complexes over $Fuk(Y)$ (again one can consider all possible deformations
$\alpha$ at the same time, which is probably more appropriate both physically
and mathematically).  Finally, one can pass to zeroth cohomology to obtain a
triangulated category $D Fuk(Y)$. As explained above, each of these steps has
a direct justification in topological open string field theory (except for the
restriction to one-sided twisted complexes, whose physical meaning is unclear
at the topological string field theory level).

\subsection{Graded Chern-Simons field theory}  

As mentioned above, Fukaya's category admits a sector obtained by considering
a topological brane $a$ together with all branes $a[n]$ obtained from it by
shifting the grade. It is interesting to ask what happens to Fukaya's category
in such a sector. By specializing the results of [\refcite{Fukaya-book}] and
[\refcite{Fukaya_mirror2}], one finds a ($\Z$-graded) $A_\infty$ algebra which
describes string field amplitudes in this sector\footnote{The fact that such
sectors should be included in Fukaya's category was discussed in 
[\refcite{sc}]
(though it was probably known to some mathematicians).  The recent paper
[\refcite{Fukaya_mirror2}] 
does include such sectors, which are necessary if one is
to obtain a triangulated category at the end of the construction.}.  In fact,
one can consider a slightly more general case, by  taking branes whose
underlying Lagrangian cycles coincide, whose gradings span the set of all
integers $n$ and allow the Chan-Paton bundle $E_n$ of each brane to have a
different rank\footnote{ As mentioned above, this corresponds to a slight
extension of the set-up of [\refcite{Fukaya_mirror2}], where only line bundles
are considered.}. (With this generalization \footnote{In fact, the papers
[\refcite{sc}, \refcite{bvf}, \refcite{gf1}, \refcite{gf2}] 
also consider a further extension, namely they allow
the connection to be complex (rather than unitary).  This can be viewed as a
way to incorporate simultaneous deformations of $L$ and $E_n$}, the relevant
sector of Fukaya's category 
need not be closed under shifts.)  The relevant construction of string
amplitudes is discussed in [\refcite{Fukaya_mirror2}]. 
Since disk instanton effects
are rather hard to compute, it is natural in first approximation to ask what
can be learned by neglecting them. This is possible close to the large radius
limit of $Y$ or when $Y=T^*L$ (since in the latter case no disk instanton
corrections are present [\refcite{Witten_CS}]). 
For simplicity, we shall also take
the B-field to vanish.

Following the approach of [\refcite{com1}, \refcite{com2}], 
the string field theory relevant
for this situation was written down in [\refcite{sc}], and turns out to be a
$\Z$-graded version of Chern-Simons field theory. This should be viewed as an
extension of Witten's original work [\refcite{Witten_CS}], obtained by adding
graded topological A-branes.  It differs markedly 
from the $\Z_2$-graded version (known
as super-Chern-Simons theory). 
Given a 3-manifold $L$, a graded Chern-Simons field theory
involves the choice of a $\Z$-graded complex superbundle ${\bf E}$, whose
degree $n$ component we denote by $E_n$. The dynamical field is a graded
superconnection ${\cal B}$ of total degree one on ${\bf E}$ in the sense of
[\refcite{Bismut_Lott}], and the action has the standard Chern-Simons form: \be
\label{gcs}
S:=\int_{L}{str\left[\frac{1}{2}{\cal B}D{\cal B}+\frac{1}{3}{\cal
B}^3\right]}~~, 
\ee where one uses graded multiplication in the space of sections of $End({\bf
E})\otimes \Lambda^* (L)$ (endowed with the total grading) 
and $str$ is the associated supertrace. We refer the
reader to [\refcite{sc}] for details.  The equations of motion of (\ref{gcs}) 
require the graded superconnection ${\cal B}$ to be flat in the sense of 
[\refcite{Bismut_Lott}].

When $L$ is a Lagrangian in $X$, this theory
naturally captures topological open string dynamics in the absence of
worldsheet instanton corrections (and the effect of open string instantons at
tree level is described by a sector of Fukaya's category as mentioned above
--- as we discuss below, this can be included as a correction to the
superpotential of [\refcite{gf2}]).  
As explained in [\refcite{sc}], fluctuations $\phi$
around a background flat graded 
superconnection ${\cal B}_0$ describe open string
states stretching between $L$ and its grade-translates $L[n]$.  To see this,
let us limit from now on to the case when 
$E_n=E$ for all $n$ (i.e. all Chan-Paton bundles
coincide) and notice that $\phi$ is a degree one section of $End({\bf
E})\otimes \Lambda^* (L)$, which can be expanded as: \be \phi=\sum_{m,n\in
\Z}{\phi_{mn}}~~, \ee where $\phi_{mn}$ are sections of 
$Hom(E_m,E_n)\otimes \Lambda^{1+m-n}L$. Then $\phi_{mn}$ describes a state of the open string stretching from
$a[m]$ to $a[n]$ (this state has worldsheet $U(1)$ charge equal to one).  The
equations of motion are: \be
\label{gcs_mc}
d\phi +\frac{1}{2}\left[\phi, \phi\right]=0~~, \ee where $d$ is the de Rham
differential on $L$ coupled to ${\cal B}_0$ and $\left[\cdot,\cdot\right]$ is
the graded commutator in the graded associative algebra ${\cal H}$ of sections
of $End({\bf E})\otimes \Lambda^* (L)$. 
Expanding $\phi$ in (\ref{gcs_mc}) shows
that $\phi_{mn}$ satisfy equations (\ref{twisted_complex}), and thus the
collection $(\phi_{mn})$ defines a twisted complex of degree one. 
Note that the twisted
complexes resulting from (\ref{gcs}) are not one-sided, since there is no
physical reason to require $\phi_{mn}=0$ if $n \leq m$. The original motivation
for studying the theory (\ref{gcs}) goes back to the announcement 
[\refcite{com2}]
and consists in understanding the physical role of twisted complexes which are
not one-sided. As we have just seen, such complexes appear naturally in the
theory (\ref{gcs}).  It is also easy to show 
[\refcite{sc}, \refcite{bvf}] that condensation
of twisted complexes gives an {\em explicit} description of extended
deformations of the pair $(L,E)$.  By this we mean that one can {\em integrate
out} finite extended deformations, namely they are explicitly represented as
twisted complexes of bundle-valued forms and maps. An explicit representation
of finite deformations is difficult to give at the level of the associated
triangulated category, since it requires one to represent the homotopy
Maurer-Cartan functor discussed in Section \ref{osft}.

The innocently-looking action (\ref{gcs}) describes a rather complicated
system, a fact which becomes apparent when one expands ${\cal B}$ in its
components.  To study its dynamics, one must at least understand how to
perform appropriate gauge-fixing. Since the theory involves higher rank forms,
this requires the full force of the BV formalism, which was applied to the
action (\ref{gcs}) in the papers [\refcite{bvf}] and [\refcite{gf2}]. The appropriate BV
action for (\ref{gcs}) was written down in [\refcite{sc}] and was analyzed in
detail in [\refcite{bvf}], where it was showed that it satisfies the classical 
master
equation. 

The general proof of the classical master equation given in [\refcite{bvf}] 
rests
on a certain {\em graded} version of the geometric formalism of BV systems due
to [\refcite{Witten_antibracket}, \refcite{Kontsevich_Schwarz}] 
(this is necessary because
the ghost number is an integer rather than $\Z_2$ valued) and makes use of the
graded supermanifolds of [\refcite{Voronov_graded}]. While this requires one to develop
some formalism, it allows for a one-line proof of the final result. An
interesting by-product of the analysis is a certain periodicity mod 6 of the
BV description under grade shifts, which 
is related to the periodicity observed in
[\refcite{Douglas_cat}].  The resulting BV action can be formulated quite
succinctly in supergeometric terms (see [\refcite{sc}, \refcite{bvf}, 
\refcite{gf2}]). Gauge fixing of
this action was performed in [\refcite{gf2}], 
and requires an infinite triangle of
antifields and auxiliary fields. The final result can be written in a form
which resembles that of usual Chern-Simons theory, though it contains
considerably more complicated dynamics.

Another result extracted in [\refcite{gf2}] is an expression for the associated
tree-level potential, which (as predicted by the 
general discussion of Section \ref{def}) 
is encoded by a series of $A_\infty$ products
describing tree-level scattering amplitudes. The perturbative 
expansion of this potential
can be written down explicitly by applying the method of [\refcite{superpot}] to
this graded set-up. In open string language, the fact that one obtains
nontrivial higher products in this manner is the effect of the so-called `open
instantons at infinity of the moduli space' pointed out in [\refcite{Witten_CS}]
and discussed (in that case) in [\refcite{Fukaya_ctg}]. The results of
[\refcite{superpot}] imply that the $A_\infty$ algebra defined by the
tree-level scattering products is cyclic with respect to a certain bilinear
form and homotopy equivalent with the differential graded algebra $({\cal H},
d, \bullet)$, where ${\cal H}$ is the space of sections of $End({\bf
E})\otimes \Lambda^*(L)$, $d$ is the de Rham differential coupled to the
background flat graded superconnection ${\cal B}_0$ and $\bullet$ is the
associative product on ${\cal H}$. One can further pass to 
a countably-generated subcomplex $C$ of 
$W^{-\infty}(End({\bf E})\otimes \Lambda^*L)$ whose components $C(m,n)$ 
are constructed as in
[\refcite{Fukaya_mirror2}] and mentioned above.  It then follows from the work of
[\refcite{Fukaya-book}, \refcite{Fukaya_mirror2}] that the effect of disk instanton corrections is
(up to an irrelevant homotopy equivalence which amounts to a change of
coordinates on the local moduli space) to deform the $A_\infty$ algebra
obtained through gauge-fixing in [\refcite{gf2}] to the $A_\infty$ algebra of
[\refcite{Fukaya-book}, \refcite{Fukaya_mirror2}] which encodes the instanton corrections; this of
course deforms the associated tree-level potential $W_{tree}$. 
With the extension to
complex graded superconnections ${\cal B}$ used in [\refcite{sc}, 
\refcite{bvf}, \refcite{gf1}, \refcite{gf2},
\refcite{superpot}], the deformed potential $W_{def}$ obtained in this manner should be
viewed as a (graded) A-model version of the D-brane superpotential of
[\refcite{Kachru_superpot}].

Since the theory is topological, the partition function of (\ref{gcs}) should
provide a topological invariant of triples $(L, {\bf E}, {\cal B})$ where 
${\cal B}$
is a flat graded superconnection on ${\bf E}$. This is a sort of analogue of
Witten's invariant, whose `quantization' by including instanton corrections
could teach us something interesting about the diagonal sector of Fukaya's
category. Unfortunately, it is currently unknown how to determine this
invariant, for example whether the surgery arguments of 
[\refcite{Witten_surgery}]
admit an extension to this graded case. However, it is possible to build a
simpler invariant, namely a graded version of the analytic torsion of Ray and
Singer (a form of Ray-Singer torsion coupled to a graded flat
superconnection). Physically, this differential invariant 
arises upon expanding the partition function of
(\ref{gcs}) around the given flat graded superconnection, and regularizing the
relevant volume factors. The necessary analysis was performed in 
[\refcite{rs}],
which extracts an explicit expression for the result in terms of a graded
version of the Ray-Singer norm and proves independence of this norm of the
choice of auxiliary 
metric data used to perform the gauge fixing. An different derivation is
given in [\refcite{gf2}] in the framework of the BV formalism. Note that the graded
Ray-Singer torsion encodes one-loop effects in the graded Chern-Simons theory,
which correspond to open string loops of the underlying topological string (in
the absence of instanton corrections). As such, the information contained in
this invariant goes beyond string tree level. It is currently
unknown how to effectively include the effect of worldsheet instanton
corrections to the graded Ray-Singer torsion. Finally, it is quite obvious
that a very similar analysis can be carried out for the holomorphic graded
Chern-Simons theory discussed in Section \ref{Bmodel}. Most of the relevant
results can be obtained simply by performing appropriate substitutions in the
final results of [\refcite{sc}, \refcite{bvf}, \refcite{gf1}, 
\refcite{gf2}]. In particular, one can write down a
graded version of holomorphic Ray-Singer torsion.

\section{$\Pi$-stability}
\label{stability}

In the paper [\refcite{DA}] (see also [\refcite{Douglas_cat}, 
\refcite{Douglas_note} and the earlier work \refcite{Douglas_stab0}]), 
M.~Douglas and P~Aspinwall
proposed a stability condition for objects of $D^b(X)$ when $X$ is a
Calabi-Yau threefold. This proposal was formulated rigorously in 
[\refcite{Bridgeland_stab}] for arbitrary triangulated
categories. Given a triangulated category ${\cal T}$, a {\em stability
  condition} [\refcite{Bridgeland_stab}] is a pair $(Z, {\cal P})$ where:

(1) $Z:K({\cal T})\rightarrow \C$ is a complex-valued linear map on the
    $K$-group $K({\cal T})$ of ${\cal T}$

(2)${\cal P}$ is a family of full subcategories ${\cal P}(\phi)$, indexed by a
real parameter $\phi$ and satisfying the following requirements:

(2a)$\phi = \frac{1}{\pi} arg Z([E])~(mod~2)$ for all $E\in {\cal P}(\phi)$ (where $[E]$ is 
the K-theory class of $E$).

(2b)${\cal P}(\phi+1)={\cal P}(\phi)[1]$ for all real $\phi$

(2c) $Hom_{\cal T}(E_1, E_2)$ vanishes if $E_1\in {\cal P}(\phi_1)$ and
$E_2\in {\cal P}(\phi_2)$ with $\phi_1>\phi_2$

(2d) For every nonzero object $E$ of ${\cal T}$ there exists a finite
descending sequence $\phi_1>\phi_2\dots >\phi_n$ and a collection of triangles
$E_j\rightarrow E_{j+1}\rightarrow A_{j+1}\rightarrow E_j[1]$ (with $j=0\dots
n-1$ and $E_n=E$) such that $A_j\in {\cal P}(\phi_j)$ for all $j$.

With this definition, one can show [\refcite{Bridgeland_stab}] that each
subcategory ${\cal P}(\phi)$ is Abelian.  The objects of ${\cal P}(\phi)$ are
called {\em semistable objects of phase $\phi$}, while the simple semistable
objects are called {\em stable}.  One also shows 
[\refcite{Bridgeland_stab}] that
the `decomposition' at point (2d) is unique for every nonzero 
object $E$. The objects
$A_j$ are called the {\em semistable factors} of $E$.

In the context of an open superstring compactification on a Calabi-Yau
threefold $X$, this data arises as follows. The triangulated category ${\cal
T}$ is the bounded derived category $D^b(X)$. The stringy Kahler moduli space
of $X$ (which arises from closed string mirror symmetry) coincides with the
complex moduli space ${\cal M}$ of the mirror $Y$ of $X$. Given a choice of
large complex structure point of $Y$ (identified with a large radius point of $X$) in ${\cal M}$, there is an associated `topological
mirror map' $m:K(D^b(X))\rightarrow H_3(Y,\Z)$ which implements open mirror
symmetry at the level of D-brane {\em charges}\footnote{This map depends on the choice of 
large radius point, which specifies a semiclassical limit of the model. There exist models (such 
as a model studied in [\refcite{arith2}]) which have more than one large radius limit, and 
where this dependence can be seen very explicitly and has pronounced physical 
consequences for the D-brane spectrum.}. Fixing a point $q$ in ${\cal
M}$, one considers the linear map $Z_q:K({\cal T})\rightarrow \C$ given by:
\be Z_q([E])=\int_{m([E])}{\Omega_q}~~, \ee where $\Omega$ is the (normalized)
holomorphic 3-form of $Y$ associated to the complex structure defined by
$q$. The quantity $Z_q([E])$ is the central charge of any B-type brane in the
K-theory class $[E]$. This naturally depends on the stringy Kahler moduli of
$X$, described by the point $q$.  The quantity $\phi$ defined at (2a) is the
real-valued grade of [\refcite{Douglas_cat}].  In particular, the worldsheet $U(1)$
charge of a superstring state stretching between branes $A,B$ of phases
$\phi_A,\phi_B$ is shifted by $\phi_B-\phi_A$.  Condition (2b) reflects the
relation between the real-valued grade $\phi$ and the integral grade used in
the associated B-twisted version of topological strings, which was discussed
above (in particular, the integral grade fixes a branch of the argument
appearing in the definition of $\phi$). Condition (2c) also follows from
conformal field theory considerations in the untwisted theory.  Condition (2d)
is related to the existence of a finite mass gap in the spectrum of BPS
states.  The objects of ${\cal P}(\phi)$ describe [\refcite{Douglas_cat}, 
\refcite{DA}] the
stable B-type branes whose central charge has phase $\phi$. The phase
specifies the particular ${\cal N}=1$ supersymmetry preserved by the
associated BPS state (i.e. the particular combination of the original ${\cal
N}=2$ generators which survives when including this state).  The
`decomposition' at (2d) above describes the decay products of a brane $E$; $E$
is semistable when $n=1$.  The mass of such a brane is given by
$m(E)=\sum_{j}{|Z(A_j)|}$, with $A_j$ its semistable factors. By the triangle
inequality, one has $m(E)\geq |Z(E)|$, with equality if $E$ is semistable;
this implements the Bogomolnyi bound.

We now present a result of [\refcite{Bridgeland_stab}], which clarifies the
relation of these notions with more standard constructions in the theory of
triangulated categories. Let us return to a triangulated category ${\cal
T}$. Remember that a {\em t-structure} on ${\cal T}$ is a full subcategory
${\cal F}$ of ${\cal T}$ such that:

(1)$ {\cal F}[1]\subset {\cal F}$

(2) For every objects $E$ in ${\cal T}$, there exists a triangle $F\rightarrow
E \rightarrow G\rightarrow F[1]$ of ${\cal T}$ with $F\in {\cal F}$ and $G\in
{\cal F}^\perp$.

Here ${\cal F}^\perp$ is the set of objects $F$ of ${\cal T}$ such that
$Hom_{\cal T}(E,F)$ vanishes for all objects $E$ of ${\cal F}$.

Given a t-structure ${\cal F}$, its {\em heart} is the full subcategory of
${\cal T}$ defined through: \be {\cal A}:={\cal F}\cap {\cal F}^\perp[1]~~;
\ee this is an Abelian category (whose short exact sequences are those
triangles of ${\cal T}$ all of whose vertices belong to ${\cal A}$). The main
use of t-structures is to identify Abelian subcategories of a triangulated
category in the manner indicated by this result.  A t-structure is called {\em
bounded} if ${\cal T}$ equals the union of ${\cal F}[m]\cap {\cal F}^\perp[n]$
over $m,n\in \Z$; such a t-structure can be recovered from its heart.

Also recall the notion of {\em slope function} on an Abelian category ${\cal
A}$ [\refcite{Rudakov}, \refcite{Bridgeland_stab}] and the associated notion of stability.  A
slope function is a linear function $Z:K({\cal A})\rightarrow \C$ such that
$\phi(E):=\frac{1}{\pi}arg Z(E)\in (0,1]$ for all nonzero objects $E$ of
${\cal A}$ (in [\refcite{Bridgeland_stab}], such slope functions are called
`centered'). Given a slope function, a nonzero 
object $E$ of ${\cal A}$ is called
{\em semistable} if $\phi(A)\leq \phi(E)$ for any nontrivial exact sequence
$0\rightarrow A \rightarrow B \rightarrow C\rightarrow 0$ in ${\cal A}$.  A
slope function satisfies the {\em Harder-Narasimhan} property if every object
of ${\cal A}$ admits an ascending filtration $0=E_0\hookrightarrow \dots 
\hookrightarrow E_n=E$ whose quotients $F_j=E_j/E_{j-1}$ are semistable and of decreasing
slope (i.e. $\phi(F_1)>\dots >\phi(F_n)$) (such a filtration is called a {\em
Harder-Narasimhan filtration}).

The following result is proved in [\refcite{Bridgeland_stab}]:

\begin{proposition}
Giving a stability condition on ${\cal T}$ amounts to giving
 a bounded t-structure, together with a slope function on its heart satisfying
 the Harder-Narasimhan property.
\end{proposition}

Given the stability condition $({\cal P}, Z)$, the t-structure is recovered as
follows. For each interval $I$ lying along the real axis, one defines ${\cal
P}(I)$ to be the extension-closure in ${\cal T}$ of the collection ${\cal
P}(\phi)$ with $\phi\in I$ (the extension closure is taken with respect to the
triangles of ${\cal T}$). Then ${\cal F}={\cal P}(0,\infty)$ is a bounded
t-structure on ${\cal T}$, whose heart equals ${\cal A}={\cal P}(0,1]$.  The
slope function on ${\cal A}$ is obtained by restricting $Z$. The semistable
objects in ${\cal A}$ defined by this slope function are the objects of ${\cal
P}(\phi)$ with $\phi\in (0,1]$, and the decompositions at (2d) above give the
Harder-Narasimhan filtrations in ${\cal A}$. The converse also follows in a
pretty obvious manner (see [\refcite{Bridgeland_stab}] for details).

This result shows that the stability condition of [\refcite{Douglas_cat}, 
\refcite{DA}] is a
natural extension of the standard stability condition on an Abelian category,
as formulated by Rudakov [\refcite{Rudakov}]. It also allows one to recover the
stable objects of ${\cal T}$, provided that one can detect the later on an
appropriate Abelian subcategory. Among these is the beautiful fact that
(under certain technical assumptions), the set of so-called {\em numerical}
stability conditions on ${\cal T}$ carries a natural topology with respect to
which every connected component is a manifold.

Let us now return to the physically interesting case (namely ${\cal T}=D^b(X)$
with $X$ a smooth Calabi-Yau threefold) and briefly recall some other aspects
discussed in [\refcite{Douglas_cat}, \refcite{DA}].  Since the central charge
$Z$ depends on the complexified stringy Kahler moduli of $X$, the subcategory
${\cal P}(\phi)$ will vary with such moduli.  This phenomenon (which the
authors of [\refcite{Douglas_cat}, \refcite{DA}] call `flow of gradings') is analogous to the
dependence of $\mu$-stability of coherent sheaves on the choice of Kahler
class.
 
A related aspect is the existence of monodromies of the periods of $Y$ around
components of the discriminant locus in ${\cal M}$, which imply that $\phi$ is
multivalued when defined on ${\cal M}$. Physical consistency requires that its
monodromy transformations should correspond to autoequivalences of $D^b(X)$,
relating the subcategories ${\cal P}(\phi)$ associated to different branches
of $\phi$. Some evidence for this conjecture was given in [\refcite{DA}], upon
using previous work of [\refcite{Seidel_Thomas}]. Further aspects of this fascinating proposal are
discussed in [\refcite{Aspinwall_D0}], which considers its physical consequences
for certain classes of $D0$-branes and what their physics may teach us in relation to the 
reconstruction results of [\refcite{BO_reconstruction}].

The proposal of [\refcite{Douglas_cat}, \refcite{DA}] amounts to a physically-motivated
definition, whose applications are currently under-explored. In
algebraic geometry, stability conditions traditionally play a role in the global
construction of moduli spaces as algebro-geometric objects (stacks, schemes or
even varieties, in a few lucky cases).  From this perspective, the deeper
mathematical role of the stability condition of [\refcite{Douglas_cat}, 
\refcite{DA}, \refcite{Bridgeland_stab}] could be found by connecting it with moduli problems. In particular, 
one would like to understand the relation of such a global study to
the local description afforded by the $A_\infty$ structure on $D^b_\infty(X)$.

\section{Some applications}
\label{applications}

\subsection{Stability, monodromies and derived equivalences}

Homological mirror symmetry predicts the existence of certain autoequivalences
of $D^b(X)$ associated with monodromy transformations around the discriminant
locus in ${\cal M}$.  This was pointed out by Kontsevich
[\refcite{Kontsevich_Rutgers}] and analyzed in detail in 
[\refcite{Seidel_Thomas}, \refcite{Seidel_Khovanov}, 
\refcite{Seidel_sequence}, \refcite{Horja_EZ}]. Here is the basic idea. It is
known [\refcite{Seidel_knotted}, \refcite{Seidel_graded}] that any Lagrangian sphere $S$ in
$Y$ defines a symplectic automorphism of $Y$ called {\em generalized Dehn
twist} along $S$ (a symplectic version of Picard-Lefschetz
transformation). Such automorphisms involve a choice of local data, but the
induced functor on $D Fuk(Y)$ is expected to be independent of such
choices and lead to an autoequivalence of $D Fuk(Y)$ 
(see [\refcite{Seidel_mutation1}, \refcite{Seidel_mutation2}] 
for a study in a local
context). By homological mirror symmetry, the mirror of $S$ should be a 
{\em spherical object} $E$ of $D^b (X)$, i.e. an
object for which $Ext^*(E)$ is concentrated in degrees zero and $dimX$, in 
which degrees it equals $\C$.  
The
Dehn twist action on $D Fuk(Y)$ should correspond to an autoequivalence
$T_E$ of $D^b(X)$ induced by $E$. 

Using the known action of Dehn twists on $H_*(Y)$ and the topological mirror
map, one finds a prediction for the action induced on $K(D^b(X))$, which
suggests an ansatz for the autoequivalence. Namely, $T_E$ should be given by:
\be
\label{TE}
T_E(F)=Cone(Hom^*(E,F)\otimes_\C E\rightarrow F)~~, \ee where the arrow is the
evaluation map. Of course, this only determines $T_E(F)$ up to a non-canonical
isomorphism, an ambiguity which is eliminated by the following precise
definition:

\begin{definition}
The {\em twist functor $T_E$} defined by an object $E$ of
$D^b(X)$ is the Fourier-Mukai transform with kernel\footnote{The definition
works for objects $E$ which are complexes of locally free sheaves and extends
correctly to $D^b(X)$.}: \be {\cal P}={\rm Cone}(\eta:E^v\boxtimes
E\rightarrow {\cal O}_\Delta)\in D^b(X\times X)~~, \ee where $\Delta$ is the
diagonal of $X\times X$, $\eta$ is the natural pairing and
$\boxtimes$ is the exterior tensor product.  
\end{definition}

Thus $T_E(F)={\bf
R}\pi_{2*}(\pi_1^* F\otimes^{\bf L} {\cal P})$, where $\pi_j$ are the two
projections on the factors of $X\times X$. Then  [\refcite{Seidel_Thomas}] 
proves
the following result, which agrees with homological mirror symmetry
expectations:

\begin{theorem}
If $E$ is a spherical object, then $T_E$ is an autoequivalence
of $D^b(X)$.
\end{theorem}

In fact, the paper [\refcite{Seidel_Thomas}] also analyzes a more general
situation, which arises for example when studying resolutions of Calabi-Yau
quotient singularities. Namely, they study so-called $A_n$-configurations of
objects in $D^b(X)$, which are expected to be mirror to similar configurations
of Lagrangians in $D Fuk(Y)$.  In this case, they prove that the associated
twist functors define a braid group action on $D^b(X)$, which is faithful for
$dim X >1$.

The beautiful results of [\refcite{Seidel_Thomas}] where discussed from a physical
perspective in [\refcite{DA}, \refcite{Aspinwall_massless}] (for the case $dim X=3$). 
The basic observation is that
the description (\ref{TE}) of the twist functor can be justified by
considering the effect of monodromies on the $\Pi$-stability condition on
$D^b(X)$. This interpretation arises when one has a spherical object $E$ which
becomes massless along a certain component of the discriminant locus. In the
toric case, the work of [\refcite{Yakov}, \refcite{arith1}, 
\refcite{arith2}, \refcite{Horja_hypergeom}] suggests
that the spherical object ${\cal O}_X$ (which describes a $D6$-brane wrapping
$X$\footnote{And whose expected mirror is a sphere in $Y$
[\refcite{SYZ}, \refcite{Gross_top1}, \refcite{Gross_top2}, 
\refcite{Gross_top3}].})  is always massless along
the principal component of the discriminant locus. This allows one to argue
[\refcite{DA}, \refcite{Aspinwall_massless}] that the associated twist functor $T_{{\cal
O}_X}$ is compatible with the $\Pi$-stability condition of [\refcite{DA}]. 
Whether
the $\Pi$-stability condition in fact {\em implies} the form (\ref{TE}) of
this monodromy action (as conjectured in [\refcite{DA}, 
\refcite{Aspinwall_massless}]) can
presumably be proved by using the rigorous definition given in
[\refcite{Bridgeland_stab}].

A far-reaching extension of the results of [\refcite{Seidel_Thomas}] was given in
[\refcite{Horja_EZ}], which constructs a large class of autoequivalences of
$D^b(X)$ induced by so-called "EZ-spherical objects".  These give a
generalization of the spherical objects of [\refcite{Seidel_Thomas}].  Such
autoequivalences are associated with a flat morphism from a smooth complete
subvariety $E$ of $X$ to another smooth subvariety $Z$ of lower dimension. In
the toric case, this situation can be realized by considering contractions
associated with various components of the discriminant locus; the principal
discriminant then corresponds to $Z= \{a~point\}$ and recovers the analysis of
[\refcite{Seidel_Thomas}].  The physical interpretation of this larger class of
autoequivalences was discussed in [\refcite{Aspinwall_massless}], which again
relates them to the effect of monodromy transformations on the $\Pi$-stability
condition.

\subsection{Autoequivalences induced by flops, worldvolume 
theories and "toric duality"}

Another mathematical result which admits a physical interpretation is due to
[\refcite{Bridgeland_flops}] (see [\refcite{BO_flops}] for earlier results):

\begin{theorem}
Let $S$ be a projective threefold with terminal singularities
and let $X, X'$ be two crepant resolutions of $S$.  Then there exists a
equivalence of triangulated categories between $D^b(X)$ and $D^b(X')$.
\end{theorem}

By results of [\refcite{Kollar}], any two such resolutions are related by a finite
chain of flops. For a flop, the result follows [\refcite{Bridgeland_flops}] by
starting with $X$ and building $X'\rightarrow S$ as a (fine) moduli space of
so-called "perverse point sheaves" constructed from the data of $X\rightarrow
S$.  Then the equivalence of the theorem is a Fourier-Mukai transform whose
kernel is constructed by using perverse sheaves.

This result implies that birational Calabi-Yau threefolds have equivalent
derived categories, a property which is physically quite natural since the
topological B-model is independent of their Kahler class.  Since the Hodge
numbers of $X$ can be extracted from the derived category, this also recovers
the invariance of Hodge numbers under flops, which was proved for general
dimension in [\refcite{Batyrev_hodge}].  For extensions of Bridgeland's result, the
reader is referred to [\refcite{flops_derived1}, \refcite{flops_derived2}].  In particular,
the paper [\refcite{flops_derived1}] generalizes this to flops between threefolds
$X,X'$ which are allowed to have terminal Gorenstein singularities.

An interesting interpretation of Bridgeland's result and its generalizations
arises upon considering (possibly partial) resolutions of Calabi-Yau quotient
singularities, which can be described in physical terms by introducing D-brane
probes transverse to the singularity [\refcite{Douglas_Greene},
\refcite{nonab}, \refcite{Morrison_Plesser}].  In this framework, an interesting situation arises for
singularities which admit multiple partial resolutions 
[\refcite{Greene_Dphases}, \refcite{BGLP}]. Then it is possible 
[\refcite{BGLP}, \refcite{BP}] that two such partial resolutions
lead to distinct but equivalent descriptions of the associated worldvolume
field theory. As discussed in [\refcite{BP}], such distinct field theory
descriptions can sometimes be related by Seiberg dualities.  A conjectural
extension of this point of view was recently proposed in
[\refcite{Douglas_Berenstein}], by making use of the derived category picture of
B-type branes. In [\refcite{Douglas_Berenstein}], it is suggested that there should
be a equivalence between the derived category of (certain ?) partial
resolutions and the derived category of the quiver which describes the
associated worldvolume theory\footnote{This should presumably follow by an
extension of the categorical McKay correspondence of 
[\refcite{Bridgeland_mckay}].}
. Based on this identification, the appropriate
generalization of Bridgeland's equivalences would translate into
autoequivalences of the Abelian category of quiver representations.  It is
known that such autoequivalences can be decomposed into tilts (this follows
from [\refcite{Rickard}] and the description of such representations as modules
over the path algebra of the quiver). On the other hand, it was argued in
[\refcite{Douglas_Berenstein}] that tilting equivalences of the the category of
quiver representations can be sometimes related to Seiberg dualities of the
associated field theory.  This is an interesting proposal which deserves
further study.  Other results on the relation between derived categories of
D-branes and supersymmetric field theory can be found in 
[\refcite{Aspinwall_sw}].

\section{Open questions}

Perhaps the most important open question in homological mirror symmetry is to
devise effective algorithms for determining the categorical mirror map. This
can be done for the case of elliptic curves 
[\refcite{Zaslow_ell}, \refcite{Polishchuk_ell1},
\refcite{Polishchuk_ell2}, \refcite{Polishchuk_ell3}, 
\refcite{Polishchuk_infty}] and higher-dimensional
Abelian varieties [\refcite{Orlov_abelian}, \refcite{Fukaya_abelian}] (see also
[\refcite{Kapustin_Orlov1}, \refcite{Kapustin_Orlov2}] for studies of complex tori).
Unfortunately, very little is known in other cases.  
For some recent progress the reader is referred to
[\refcite{Kontsevich_Soibelman}, \refcite{Fukaya_fibrations}].  A related problem is to find
effective methods for computing Fukaya's category. Some ideas in this
direction were recently proposed in [\refcite{Seidel_exact}]. Though an impressive
mathematical `tour de force', the construction of Fukaya's category is
somewhat unsatisfactory both from a physical and mathematical
perspective. Physically, the topological string field theory description of
obstructions used in Fukaya's work seems to be exceedingly abstract, and it is
hard to see how one could ever compute them through such methods. One may hope
that the alternative point of view provided by graded Chern-Simons theories can
be more fruitful from a physical perspective. Mathematically, the construction
of [\refcite{Fukaya-book}, \refcite{Fukaya_mirror2}] is so complicated that one is perhaps
better off viewing it as an existence proof. As such, it may be useful to
attempt to characterize Fukaya's category (up to homotopy equivalence) through
some abstract conditions, and use those in order to prove results about its
behavior under various geometric operations. This may allow one to compute
Fukaya's category by reduction to simpler cases.

A question of great physical importance is to understand the
extension of homological mirror symmetry to the case of mixed and
non-geometric phases, and to extend the results of [\refcite{topchange},
\refcite{Witten_topchange}] to the open string level. Some recent work along these lines 
was carried out in [\refcite{Kapustin_sings}, \refcite{Orlov_sings}], and
this entire subject deserves much more investigation. As in the closed
string case, one can shed some light on this problem 
[\refcite{Hori}, \refcite{Hori_lsm}] by
using an open version of Witten's linear sigma models
[\refcite{Witten_topchange}], though the categorical aspects of the linear
sigma model construction have not been thoroughly investigated.
Finally, it would be of great physical interest to combine such an
analysis with the stability proposal of [\refcite{DA}, 
\refcite{Bridgeland_stab}] in
order to extract a more complete description of D-brane physics on
Calabi-Yau manifolds.

\section*{Acknowledgments}

I thank M. Kapranov, A. Klemm, 
W.~Lerche, M.~Rocek, J.~Stasheff, D.~
Sullivan, H.~Verlinde, K.~Hori, A.~Kapustin and S.~Popescu 
for interest in my work. 
I also thank the editors of IJMPA for their 
patience during the long gestation period of this article. 
This work was supported by DFG grant KL1070/2-1.


\begin{thebibliography}{100}
\bibitem{Aspinwall_Lawrence}{P. Aspinwall, D. A. Lawrence, {\em Derived
      categories and zero-brane stability}, JHEP {\bf 0108} (2001) 004; 
hep-th/0104147.}
\bibitem{Aspinwall_D0}{P.~S.~Aspinwall, {\em A Point's Point of View of 
Stringy Geometry}, JHEP {\bf 0301} (2003) 002, hep-th/0203111. }
\bibitem{Aspinwall_massless}{P. S. Aspinwall, R. P. Horja, R. L. Karp, 
{\em Massless D-Branes on Calabi-Yau Threefolds and Monodromy}, 
hep-th/0209161.}
\bibitem{topchange}{ P. S. Aspinwall, B. R. Greene, D.R. Morrison, 
{\em Calabi-Yau Moduli Space, Mirror Manifolds and Spacetime Topology 
Change in String Theory}, Nucl.Phys. {\bf B416} 
(1994) 414-480;  hep-th/9309097.}
\bibitem{DA}{P.~S.~Aspinwall, M.~R.~Douglas, 
{\em D-Brane Stability and Monodromy}, JHEP {\bf 0205} (2002) 031, 
hep-th/0110071.}
\bibitem{Aspinwall_sw}{P.~S.~Aspinwall, R.~L.~Karp, 
{\em Solitons in Seiberg-Witten Theory and D-branes in
      the Derived Category}, hep-th/0211121.}
\bibitem{BGLP}{C.~Beasley, B.~R. Greene, C. I. Lazaroiu, M. R. Plesser,
{\em D3-branes on partial resolutions of abelian quotient singularities of
  Calabi-Yau threefolds}, Nucl. Phys. {\bf B566} 
(2000) 599-640, hep-th/9907186.}
\bibitem{BP}{C.~E.~Beasley, M.~R.~Plesser, 
{\em  Toric Duality Is Seiberg Duality}, JHEP {\bf 0112} (2001) 001, 
hep-th/0109053.}
\bibitem{BK}{A.~Bondal, M.~M.~Kapranov, 
{\em Enhanced triangulated categories}, Math. USSR Sbornik 
{\bf 70}(1991)1, 93.}
\bibitem{BO_reconstruction}{A.~Bondal, D.~Orlov, {\em Reconstruction of a
      variety from the derived category and groups of autoequivalences}, 
 Compositio Math. {\bf 125} (2001) 3, 327--344, alg-geom/9712029.} 
\bibitem{BO_flops}{A.~I.~Bondal, D.~O.~Orlov, 
{\em Semiorthogonal decomposition for algebraic varieties}, math.AG/9506012.}
\bibitem{Batyrev_hodge}{V.~V.~Batyrev, {\em Birational Calabi--Yau n-folds
      have equal Betti numbers},  in {\em New trends in algebraic geometry},
L.M.S. Lecture Notes Series {\bf 264}, Cambridge U. P. (1999)1-11, 
alg-geom/9710020.}
\bibitem{Bismut_Lott}{J.~M.~Bismut and J.~Lott, {\em Flat vector
      bundles, direct images and higher analytic torsion}, J. Amer.
    Math Soc {\bf 8} (1992) 291.}  
\bibitem{Bridgeland_stab}{T.~Bridgeland, {\em  Stability conditions on 
triangulated categories}, math.AG/0212237.}
\bibitem{Bridgeland_flops}{T.~Bridgeland, {\em Flops and derived categories}, 
 Invent. Math.  {\bf 147}  (2002) 3, 613--632,  math.AG/0009053.}
\bibitem{Bridgeland_mckay}{T.~Bridgeland, A.~King, M.~Reid, 
{\em Mukai implies McKay: the McKay correspondence as an equivalence of 
derived categories},  J. Amer. Math. Soc.  {\bf 14} (2001) 3, 535--554;  
math.AG/9908027.}
\bibitem{flops_derived1}{J-C.~Chen, {\em Flops and Equivalences of derived 
Categories for Threefolds with only Gorenstein Singularities};
math.AG/0202005.}
\bibitem{Douglas_Greene}{
M.~R.~Douglas, B.~R. Greene, D.~R.~Morrison, 
{\em Orbifold Resolution by D-Branes}; Nucl. Phys. {\bf B506} (1997) 84-106, 
hep-th/9704151.}
\bibitem{Douglas_stab0}{
M.~R.~Douglas, B.~Fiol, C.~R\"omelsberger, 
{\em Stability and BPS branes}, hep-th/0002037.}
\bibitem{Douglas_cat}{ M.~R.~Douglas, {\em D-branes, Categories and N=1 
Supersymmetry},  J. Math. Phys. {\bf 42} (2001) 2818-2843; hep-th/0011017.}
\bibitem{Douglas_note}{ M.~R.~Douglas, {\em Dirichlet branes, homological 
mirror symmetry, and stability};  math.AG/0207021.}
\bibitem{Douglas_Berenstein}{
D.~Berenstein, M.~R.~Douglas, 
{\em Seiberg Duality for Quiver Gauge Theories}; hep-th/0207027.}
\bibitem{Diac}{D-E.~Diaconescu, {\em  Enhanced D-Brane Categories from 
String Field Theory}, JHEP {\bf 0106} (2001) 016; hep-th/0104200.}
\bibitem{Drinfeld}{V.~Drinfeld, {\em DG quotients of DG categories}, 
 math.KT/0210114.}
\bibitem{Floer1}{A.~Floer, 
{\em Morse theory for Lagrangian intersections},
      J. Diff. Geom. {\bf 28} (1988) 513-547.}
\bibitem{Floer2}{A.~Floer, {\em Witten's complex and infinite-dimensional 
Morse theory}, J. Diff. Geom {\bf 30} (1989) 207-221.}
\bibitem{Fukaya_fibrations}{K.~Fukaya, 
{\em Floer homology for families - report
      of a project in progress}, available at 
$http\:\/\/www.kusm.kyoto-u.ac.jp/~fukaya/fukaya.html$.}
\bibitem{Fukaya_infty}{K.~Fukaya, {\em Morse homotopy, $A_\infty$ categories 
and Floer homologies}, in Proceedings of the 1993 Garc Workshop 
on Geometry and Topology, Lecture Notes series, vol 18, pages 1-102, Seoul
National Univ. 1993.}
\bibitem{Fukaya-book}{K.~Fukaya, Y.~G.~Oh, H.~Ohta, K.~Ono, {\em Lagrangian 
intersection Floer theory --- anomaly and obstruction} preprint (2000),
available at $http\:\/\/www.kusm.kyoto-u.ac.jp/~fukaya/fukaya.html$ }
\bibitem{Fukaya_rev}{K.~Fukaya, 
{\em Deformation theory, homological algebra and mirror symmetry}, 
in {\em  
Geometry and physics of branes} (Como, 2001),  121--209, 
Ser. High Energy Phys. Cosmol. Gravit., IOP, Bristol 2003.}
\bibitem{Fukaya_ctg}{K.~Fukaya, Y.~G.~Oh, {\em Zero-loop open strings in the 
cotangent bundle and Morse homotopy}, Asian. J. Math {\bf 1} (1997) 99-180.}
\bibitem{Fukaya_mirror1}{K.~Fukaya, {\em Floer homology and mirror symmetry 
I},  in {\em Winter School on Mirror Symmetry, Vector Bundles and Lagrangian 
Submanifolds} (Cambridge, MA, 1999),  15--43, AMS/IP Stud. Adv. Math., 
23, Amer. Math. Soc., Providence, RI, 2001.}
\bibitem{Fukaya_mirror2}{K.~Fukaya, {\em Floer homology and mirror symmetry 
II},  in {\em Minimal surfaces, geometric analysis and symplectic geometry} 
(Baltimore, MD, 1999),  31--127, Adv. Stud. Pure Math. {\bf 34}, 
Math. Soc. Japan, Tokyo 2002}
\bibitem{Fukaya_abelian}{K.~Fukaya, 
{\em Mirror symmetry for abelian variety and multi theta functions} (1998), 
 J. Algebraic Geom. {\bf 11} (2002)3, 393--512.}
\bibitem{Gaberdiel}{  M.~R.~Gaberdiel, B.~Zwiebach, 
{\em Tensor Constructions of Open String Theories I: Foundations}, 
 Nucl.Phys. {\bf B505} (1997) 569-624, hep-th/9705038.}
\bibitem{Yakov}{ B.~R. Greene, Y.~Kanter, {\em 
Small Volumes in Compactified String Theory}, Nucl. Phys. {\bf B497} 
(1997) 127-145; hep-th/9612181.}
\bibitem{arith1}{ B.~R.~Greene, C.~I.~Lazaroiu, 
{\em Collapsing D-Branes in Calabi-Yau Moduli Space}, Nucl. Phys. {\bf B604} 
(2001) 181-255; hep-th/0001025.}
\bibitem{nonab}{B.~R.~Greene, C. I. Lazaroiu, M.~Raugas, 
{\em D-branes on Nonabelian Threefold Quotient Singularities},
Nucl. Phys. {\bf B553} (1999) 711-749; hep-th/9811201.} 
\bibitem{Greene_Dphases}{ B.~R.~Greene, 
{\em D-Brane Topology Changing Transitions},  Nucl.Phys. 
{\bf B525} (1998) 284-296; hep-th/9711124.}
\bibitem{Gross_top1}{ M.~Gross,
{\em Special Lagrangian Fibrations I: Topology}, in {\em 
Integrable systems and algebraic geometry} (Kobe/Kyoto, 1997),  156--193, 
World Sci. Publishing, River Edge, NJ, 1998; alg-geom/9710006.} 
\bibitem{Gross_top2}{M.~Gross, {\em Special Lagrangian Fibrations II: 
Geometry}, in {\em 
Surveys in differential geometry: differential geometry inspired
by string theory},  341--403, Surv. Differ. Geom., 5, Int. Press, Boston, MA, 
1999; math.AG/9809072.}
\bibitem{Gross_top3}{M.~Gross, {\em Topological Mirror Symmetry},  
Invent. Math.  {\bf 144} (2001) 1, 75--137; math.AG/9909015 }
\bibitem{Hori}{K.~Hori, A.~Iqbal, C.~Vafa, 
{\em D-branes and mirror symmetry}, hep-th/0005247.}
\bibitem{Hori_lsm}{K.~Hori, {\em Linear models of supersymmetric D-branes}, 
hep-th/0012179.}
\bibitem{Horja_EZ}{R.~P.~Horja, {\em Derived category automorphisms from 
mirror symmetry}, math.AG/0103231.}
\bibitem{Horja_hypergeom}{R.~P.~Horja, {\em  
Hypergeometric functions and mirror symmetry in toric varieties}, 
math.AG/9912109.}
\bibitem{Kachru_superpot}{
S.~Kachru, S.~Katz, A.~Lawrence, J.~McGreevy, 
{\em Open string instantons and superpotentials}, 
Phys.Rev. {\bf D62} (2000) 026001; hep-th/9912151.}
\bibitem{Kadeishvili}{T.~D.~Kadeishvili, {\em The algebraic structure in the 
homology of an $A_\infty$ algebra}, Soobshch. Akad. Nauk. Gruzin, SSR, 
{\bf 108} (2) (1983) 249-252.}
\bibitem{Kajiura}{H.~Kajiura, {\em Homotopy Algebra Morphism and Geometry 
of Classical String Field Theory}, Nucl. Phys. {\bf B630} (2002) 361-432;
hep-th/0112228.}
\bibitem{Kapustin_sings}{A.~Kapustin, Yi Li, {\em D-Branes in Landau-Ginzburg 
Models and Algebraic Geometry}, hep-th/0210296.}
\bibitem{Kapustin_Orlov1}{A.~Kapustin, D.~Orlov,
{\em Vertex Algebras, Mirror Symmetry, And D-Branes: The Case Of Complex
  Tori}, Commun. Math. Phys. {\bf 233} (2003) 79-136; hep-th/0010293.}
\bibitem{Kapustin_Orlov2}{A.~Kapustin, D.~Orlov, 
{\em Remarks on A-branes, Mirror Symmetry, and the Fukaya category}, 
hep-th/0109098.}
\bibitem{Keller_dg}{B.~Keller, {\em Deriving DG categories}, 
Ann. Scient. Ecole Normale Sup, $4^e$ serie, t. {\bf 27} (1994) 63-102.}
\bibitem{Keller_intro}{ B.~Keller, 
{\em Introduction to A-infinity algebras and modules}, 
Homology Homotopy Appl.  {\bf 3}(2001)1, 1--35, math.RA/9910179; 
{\em Addendum to: "Introduction to $A$-infinity algebras and modules"}, 
Homology Homotopy Appl.  {\bf 4}  (2002),  no. 1, 25--28.}
\bibitem{Kollar}{J.~Kollar, {\em Flops}, Nagoya Math. J {\bf 113}(1989)
    15-36.}
\bibitem{Kontsevich_Schwarz}{
M. Alexandrov, M. Kontsevich, A. Schwarz, O. Zaboronsky, 
{\em The Geometry of the Master Equation and Topological Quantum Field
Theory}, Int.J.Mod.Phys. {\bf A12} (1997) 1405-1430; hep-th/9502010.}
\bibitem{Kontsevich0}{M.~Kontsevich, {\em Homological Algebra of Mirror
Symmetry},
{\em Proceedings of the International Congress of Mathematicians}, 
Vol. 1, 2 (Zurich, 1994)120--139, Birkhäuser, Basel 1995;  alg-geom/9411018. }
\bibitem{Kontsevich_Rutgers}{M.~Kontsevich, 
{\em Lecture at Rutgers university} (1996), unpublished.}
\bibitem{Kontsevich_Soibelman}{ M.~Kontsevich, Y.~Soibelman, 
{\em Homological mirror symmetry and torus fibrations}, 
in {\em Symplectic geometry and mirror symmetry} (Seoul, 2000),  
203--263, World Sci. Publishing, River Edge, NJ, 2001; math.SG/0011041.}
\bibitem{CIL_sft}{C.~I.~Lazaroiu, {\em Talk at ITP Stony Brook} (2000),
unpublished.}
\bibitem{arith2}{ C.~I.~Lazaroiu, {\em 
Collapsing D-branes in one-parameter models and small/large radius duality}, 
Nucl. Phys. {\bf B605} (2001) 159-191;  hep-th/0002004.}
\bibitem{top}{ C.~I.~Lazaroiu, {\em On the structure of open-closed
      topological field theory in two dimensions}, 
Nucl.Phys. {\bf B603} (2001) 497-530; hep-th/0010269.}
\bibitem{com1}{ C.~I.~Lazaroiu, {\em Generalized complexes and string field 
theory },   JHEP {\bf 0106} (2001) 052; hep-th/0102122}
\bibitem{com2}{ C. I. Lazaroiu, {\em Unitarity, D-brane dynamics and D-brane
      categories}, JHEP {\bf 0112} (2001) 031; hep-th/0102183.}
\bibitem{sc}{ C.~I.~Lazaroiu, {\em Graded Lagrangians, exotic topological
      D-branes and enhanced triangulated categories}, JHEP {\bf 0106} 
(2001) 064; hep-th/0105063 }
\bibitem{superpot}{ C.~I.~Lazaroiu, {\em String field theory and brane 
superpotentials}, JHEP {\bf 0110} (2001) 018; hep-th/0107162.}
\bibitem{bvf}{ C.~I.~Lazaroiu,  R. Roiban, D. Vaman,  
{\em Graded Chern-Simons field theory and graded topological D-branes}, JHEP
{\bf 0204} (2002) 023; hep-th/0107063.}
\bibitem{gf1}{ C.~I.~Lazaroiu, R.~Roiban {\em Holomorphic potentials for 
graded D-branes}, JHEP {\bf 0202} (2002) 038; hep-th/0110288. }
\bibitem{gf2}{ C.~I.~Lazaroiu, R.~Roiban, 
{\em Gauge-fixing, semiclassical approximation and potentials for 
graded Chern-Simons theories},  JHEP {\bf 0203} (2002) 022; hep-th/0112029.}
\bibitem{rs}{ C.~I.~Lazaroiu, {\em An analytic torsion for graded D-branes}, 
JHEP {\bf 0209} (2002) 023; hep-th/0111239.}
\bibitem{Lefevre}{K.~Lefevre-Hasegawa, 
{\em  Sur les $A_\infty$ categories}, Ph.D. thesis. Available at 
http://www.math.jussieu.fr/~lefevre/publ.html.}
\bibitem{Moore_top}{G.~Moore and G.~Segal, {\em unpublished.} Available at 
$http://online.kitp.ucsb.edu/online/mp01/moore1/ and /moore2$.}
\bibitem{moore_K}{G.~Moore, {\em Some Comments on Branes, G-flux, and 
K-theory}, Int. J. Mod. Phys. {\bf A16} (2001) 936-944, hep-th/0012007.}
\bibitem{Morrison_Plesser}{ D.~R.~Morrison, M.~R.~Plesser,
{\em Non-Spherical Horizons, I}, Adv. Theor. Math. Phys. {\bf 3}
(1999) 1-81, hep-th/9810201.}
\bibitem{flops_derived2}{Y.~Namikawa, {\em Mukai flops and derived 
categories}, math.AG/0203287. }
\bibitem{Oh}{Y.-G.~Oh, {\em Floer cohomology of Lagrangian
intersections and pseudo-holomorphic disks I, II}, Comm. Pure
Appl. Math. {\bf 46}(1993), 949-994 and 995-1012; {\em Floer
cohomology of Lagrangian intersections and pseudo-holomorphic disks
III}, Floer memorial volume, Birkhauser, Basel, 1995, 555-573.}
\bibitem{Orlov_abelian}{ V. Golyshev, V. Lunts, D. Orlov, 
{\em Mirror symmetry for abelian varieties},  J. Algebraic Geom.  
{\bf 10}(2001) 3, 433--496; math.AG/9812003.}
\bibitem{Orlov_sings}{D.~Orlov, {\em Triangulated categories of singularities
      and D-branes in Landau-Ginzburg models}, math.AG/0302304.}
\bibitem{Zaslow_ell}
{A.~Polishchuk, E.~Zaslow, {\em Categorical mirror symmetry:
the elliptic curve}, Adv. Theor. Math. Phys. {\bf 2} (1998) 443-470.}
\bibitem{Polishchuk_ell1}{A.~Polishchuk,
{\em Massey and Fukaya products on elliptic curves}, 
 Adv. Theor. Math. Phys.  {\bf 4} (2000) 6, 1187--1207; math.AG/9803017.}
\bibitem{Polishchuk_ell2}{A.~Polishchuk, {\em  $A_{\infty}$-structures on an 
elliptic curve},  math.AG/0001048.}
\bibitem{Polishchuk_ell3}{A.~Polishchuk, {\em  Indefinite theta series of 
signature (1,1) from the point of view of homological mirror symmetry},  
math.AG/0003076.}
\bibitem{Polishchuk_infty}{A.~Polishchuk, {\em Homological mirror symmetry 
with higher products},  in {\em 
Winter School on Mirror Symmetry, Vector Bundles and
Lagrangian Submanifolds} (Cambridge, MA, 1999),  247--259, AMS/IP
Stud. Adv. Math., 23, Amer. Math. Soc., Providence, RI, 2001; 
math.AG/9901025.} 
\bibitem{Rickard}{J. Rickard, {\em Morita theory for derived categories}, 
J. London Math. Soc. {\bf 39} (1989) 436-456.}
\bibitem{Rudakov}{A.~Rudakov, {\em Stability for an abelian category}, 
 J. Algebra  {\bf 197}  (1997),  no. 1, 231--245. }
\bibitem{SYZ}{A.~Strominger, S.~T.~Yau, E.~Zaslow, {\em Mirror Symmetry is
      T-Duality}, Nucl. Phys. {\bf B479} (1996) 243-259; hep-th/9606040.}
\bibitem{Seidel_knotted}
{P.~Seidel, {\em Lagrangian two-spheres can be
      symplectically knotted}, J. Differential Geom. {\bf 52} (1999), 145-171, 
math.DG/9803083.}
\bibitem{Seidel_graded}{P.~Seidel, {\em  Graded Lagrangian submanifolds}, 
Bull. Soc. Math. France {\bf 128} (2000), 103-149, math.SG/9903049.}
\bibitem{Seidel_Thomas}
{P.~Seidel, R.~Thomas, {\em Braid group actions on derived 
categories of coherent sheaves}, Duke. Math. J. {\bf 108}(1)(2001)37-108.}
\bibitem{Seidel_Khovanov}{M.~Khovanov; P.~Seidel,
{\em Quivers, Floer cohomology, and braid group actions}, 
J. Amer. Math. Soc.  {\bf 15} (2002) 1, 203--271.}
\bibitem{Seidel_sequence}{P.~Seidel, {\em 
A long exact sequence for symplectic Floer cohomology}, math.SG/0105186.} 
\bibitem{Seidel_exact}{P.~Seidel, {\em Fukaya categories and deformations}, 
math.SG/0206155.}
\bibitem{Seidel_mutation1}{P.~Seidel,  {\em Vanishing cycles and mutation}, 
in {\em European Congress of Mathematics}, Vol. II (Barcelona, 2000),  
65--85, Progr. Math., 202, Birkhäuser, Basel, 2001;  math.SG/0007115.}
\bibitem{Seidel_mutation2}{P. Seidel,  {\em More about vanishing cycles and 
mutation},  in {\em Symplectic geometry and mirror symmetry} (Seoul, 2000), 
429--465, World Sci. Publishing, River Edge, NJ, 2001; math.SG/0010032.}
\bibitem{Sharpe}{A.~Caldararu, S.~Katz and E.~Sharpe,
{\em D-branes, B fields, and Ext groups},hep-th/0302099.}
\bibitem{Thomas_Yau}{ R.~P.~Thomas, S.-T. Yau, 
{\em Special Lagrangians, stable bundles and mean curvature flow}, 
Communications in Analysis and Geometry {\bf 10}, 
1075-1113, 2002; math.DG/0104197}
\bibitem{Thomas_stab}{ R. P. Thomas, {\em 
Stability conditions and the braid group}, math.AG/0212214.}
\bibitem{Voronov}{A.~A.~Voronov, 
{\em Topological field theories, string backgrounds and homotopy algebras}, 
hep-th/9401023, Adv. Appl. Clifford Algebras 4 (1994), Suppl. 1, 167--178. } 
\bibitem{Voronov_graded}{T.~Voronov, {\em 
Graded manifolds and Drinfeld doubles for Lie bialgebroids}, 
math.DG/0105237.}
\bibitem{Witten_topchange}{E.~Witten, {\em  Phases of $N=2$ Theories In Two 
Dimensions}, Nucl.Phys. {\bf B403} (1993) 159-222; hep-th/9301042.}
\bibitem{Witten_Zwiebach}{E.~Witten, B.~Zwiebach, 
{\em Algebraic Structures and Differential Geometry in 2D String Theory}
Nucl.Phys. {\bf B377} (1992) 55-112; hep-th/9201056.}
\bibitem{Witten_nlsm}{
E.~Witten, {\em Topological sigma models}, Commun. Math. Phys. 
{\bf 118} (1988),411.}
\bibitem{Witten_mirror}{E.~Witten, 
{\em Mirror manifolds and topological field theory}, 
Essays on mirror manifolds, 120--158, Internat. Press, 
Hong Kong, 1992; hep-th/9112056.}
\bibitem{Witten_surgery}{E.~Witten, {\em Quantum field theory and the 
Jones polynomial}, Commun Math Phys {\bf 121} (1989) 351.}
\bibitem{Witten_antibracket}{E.~Witten, {\em A note on the antibracket 
formalism}, Mod. Phys. Lett. {\bf A5} (1990) 487.}
\bibitem{Witten_SFT}{E.~Witten, {\em Noncommutative geometry and string 
field theory}, Nucl. Phys {\bf B268} (1986) 253.}
\bibitem{Witten_CS}{
E.~Witten, {\em Chern-Simons gauge theory as a string theory}, 
The Floer memorial volume, 637--678, Progr. Math. {\bf 133}, Birkhauser, Basel,
1995; hep-th/9207094.}
\bibitem{Zwiebach_oc}{B.~Zwiebach, {\em Oriented Open-Closed String Theory 
Revisited}, Annals Phys. {\bf 267} (1998) 193-248; hep-th/9705241.}
\end{thebibliography}
\end{document}